\documentclass[twocolumn]{aastex631}

\newcommand{\mbh}{$M_{\rm BH}$}

\newcommand{\msun}{$M_{\odot}$}

\shorttitle{The Broad Line Region of NGC\,3783}
\shortauthors{Bentz et al.}

\begin{document}

\title{A Detailed View of the Broad Line Region in NGC\,3783 \linebreak 
from Velocity-Resolved Reverberation Mapping}


\author[0000-0002-2816-5398]{Misty C.\ Bentz}
\affiliation{Department of Physics and Astronomy,
		 Georgia State University,
		 Atlanta, GA 30303, USA}
\email{bentz@astro.gsu.edu}

\author[0000-0002-4645-6578]{Peter R.\ Williams}
\affiliation{Department of Physics and Astronomy, 
    University of California, 
    Los Angeles, CA 90095, USA}

\author[0000-0001-6279-0552]{Rachel Street}
\affiliation{LCOGT, 6740 Cortona Drive, Suite 102, 
        Goleta, CA 93117, USA}

\author[0000-0003-0017-349X]{Christopher A.\ Onken}
\affiliation{Research School of Astronomy and Astrophysics, 
        Australian National University, 
        Canberra, ACT 2611, Australia}

\author[0000-0002-6257-2341]{Monica Valluri}
\affiliation{Department of Astronomy,
         University of Michigan,
         Ann Arbor, MI, 48104, USA}

\author[0000-0002-8460-0390]{Tommaso Treu}
\altaffiliation{Packard Fellow}
\affiliation{Department of Physics and Astronomy, 
    University of California, 
    Los Angeles, CA 90095, USA}

\begin{abstract}
We have modeled the full velocity-resolved reverberation response of the H$\beta$ and \ion{He}{2} optical broad emission lines in NGC\,3783 to constrain the geometry and kinematics of the low-ionization and high-ionization broad line region.  The geometry is found to be a thick disk that is nearly face on, inclined at $\sim18\degr$ to our line of sight, and exhibiting clear ionization stratification, with an extended H$\beta$-emitting region ($r_{\rm median}=10.07^{+1.10}_{-1.12}$\,light~days) and a more compact and centrally-located \ion{He}{2}-emitting region ($r_{\rm median}=1.33^{+0.34}_{-0.42}$\,light~days). In the H$\beta$-emitting region, the kinematics are dominated by near-circular Keplerian orbits, but with $\sim 40$\% of the orbits inflowing.  The more compact \ion{He}{2}-emitting region, on the other hand, appears to be dominated by outflowing orbits. The black hole mass is constrained to be \mbh$=2.82^{+1.55}_{-0.63}\times10^7$\,\msun, which is consistent with the simple reverberation constraint on the mass based on a mean time delay, line width, and scale factor of $\langle f \rangle=4.82$.  The difference in kinematics between the H$\beta$- and \ion{He}{2}-emitting regions of the BLR is intriguing given the recent history of large changes in the ionizing luminosity of NGC\,3783 and evidence for possible changes in the BLR structure as a result.

\end{abstract}



\section{Introduction} 

Black holes continue to capture our imaginations centuries after the concept was first recorded in a letter written by a country clergyman \citep{michell1784}. Today, not only are black holes a recurrent feature in science fiction, but they have become securely ensconced in science fact.  We now know that supermassive (\mbh$=10^5-10^{10}$\,\msun) black holes exist, that they inhabit the centers of most (all?) massive galaxies, and that their masses scale with several measurable properties of their host galaxies, including the bulge stellar velocity dispersion and bulge luminosity (e.g., \citealt{magorrian98,ferrarese00,gebhardt00,gultekin09,kormendy13}). 

Only a handful of methods are able to directly constrain the mass of a central, supermassive black hole through its gravitational effects on luminous matter.  In the case of the Milky Way, astrometric monitoring of individual stars in the central few parsecs has resulted in a constraint on the mass of Sagittarius A* of \mbh$=(4.1\pm0.6)\times10^6$\,\msun \citep{ghez00,genzel00,ghez08}, while relativistic modeling of the emission from gas just outside the event horizon has constrained the mass of P\={o}wehi, the central black hole in M87, to \mbh$=(6.5\pm0.7)\times10^9$\,\msun \citep{eht19}.  Most other galaxies are not able to be studied with similar methods because we lack the necessary spatial resolution.  However, many nearby galaxies ($D\lesssim100$\,Mpc) may still be studied through spatially-resolved observations of the bulk nuclear gas or stellar kinematics on scales of $\sim$tens of parsecs (e.g., \citealt{gultekin09,kormendy13}).  Reverberation mapping is notable among black hole mass measurement techniques because it relies on time resolution rather than angular resolution.  By monitoring the spectrophotometric variability of an active galactic nucleus (AGN), the black hole mass, among other properties, may be constrained for a local Seyfert or a distant quasar (for a recent review, see \citealt{cackett21}).

Reverberation mapping makes use of the response of photoionized gas in the broad emission-line region (BLR) to variations in the continuum luminosity, a technique that was first proposed by \citet{bahcall72}.  As it is generally implemented, reverberation mapping constrains an average responsivity-weighted radius for the BLR in an AGN.  Combining the radius with a measure of the line-of-sight velocity of the BLR gas via the virial theorem constrains \mbh\ \citep{peterson99,peterson00a}, modulo a scale factor that accounts for the generally unknown BLR geometry and kinematics (e.g., \citealt{onken04,park12,grier13,batiste17b}).  However, high quality spectrophotometric monitoring data contain information about the gas response as a function of line-of-sight velocity, thus providing constraints on the emissivity and position of photoionized gas in a spatially-unresolved source \citep{blandford82}.  Velocity-resolved reverberation mapping, as it has come to be known, is thus able to directly constrain the BLR geometry and the black hole mass, thus avoiding the need to apply a scale factor.

The analysis of velocity-resolved reverberation mapping data can be approached as an ill-posed inverse problem, in which the goal is to recover the transfer function that describes the time delay distribution as a function of velocity across a broad emission line (e.g., \citealt{horne94,skielboe15,anderson21}).  Or it can be approached through direct modeling, in which a framework of fully self-consistent models is built and an exploration of the available parameter space yields the family of models that best match the observational constraints (e.g., \citealt{pancoast11}).  Direct modeling has the advantage that it is relatively simple to interpret the results, however its ability to match complicated data sets is limited by the phenomenology that is included and how it is parametrized.  Recovery of the transfer function, on the other hand, takes advantage of the full range of details present in the observations but is nontrivial to interpret. 

While the promise of velocity-resolved reverberation mapping has long been understood, it was only within the last decade or so that improvements in the quality of reverberation mapping data (e.g., \citealt{bentz08,bentz09c,denney09c,grier12b}) have finally allowed the BLR structure and kinematics to be explored in detail for a handful of AGNs \citep{pancoast14b,grier17,williams18,williams20}.  In general, direct modeling has found many similarities across objects, although the exact details vary: the low-ionization BLR is arranged in a thick disk-like structure at low to moderate inclination to our line of sight, and with kinematics that are dominated by near-circular Keplerian orbits but with a contribution from inflow (although \citet{williams18} find evidence for outflow, rather than inflow, in some of their sample).  The high-ionization BLR is less well studied, and \citet{williams20} find several key differences in not just the kinematics but also the geometry of the low- and high-ionization BLR gas in NGC\,5548.  Studies that have focused on the recovery of the transfer function have generally drawn similar conclusions about the BLR structure and kinematics \citep{bentz10b,horne21}.  A key finding of all these studies is that the black hole masses derived from a more simplistic reverberation analysis, involving a mean time delay and line width and an adopted scale factor of $\langle f \rangle \approx 5$, are generally in good agreement within their uncertainties with the masses derived from modeling.  As expected, the largest differences are generally found for those AGNs where direct modeling derives an inclination of the BLR that is $\lesssim 15^{\circ}$ to our line of sight (cf.\ Figure~14 of \citealt{williams18}).  Very low inclinations result in small observed line-of-sight velocities, which bias the simplistic mass estimates to low values. 

We recently conducted a new reverberation mapping program focusing on the bright Southern Seyfert, NGC\,3783, with the intent of improving the constraints on the black hole mass.  A nearly face-on barred spiral galaxy at $z=0.0097$, NGC\,3783 is one of the most well-studied AGNs in the sky.  It is one of a few Seyfert 1s that may be studied in detail with VLT GRAVITY observations on spatial scales that resolve the dust torus and outer broad line region \citep{gravity21}, thus it is a critical target for informing our understanding of both feeding and feedback.  Furthermore, NGC\,3783 is also one of a small number of Seyfert 1 galaxies that are near enough to allow a reverberation-based mass to be directly compared with masses constrained through dynamical methods.  The comparison of reverberation and dynamical masses is the only independent check that we can use to investigate the reliability of the entire black hole mass scale that we currently apply across cosmic history, an important point given the different systematic biases that are inherent in each black hole mass measurement technique.

An initial assessment of the monitoring data constrained a reverberation-based black hole mass of $M_{\rm BH} =(2.3\pm0.4)\times 10^7$\,M$_{\odot}$ \citep{bentz21a}, assuming a scale factor of $\langle f \rangle = 4.82$ \citep{batiste17b}.  However, variations in the time delay as a function of velocity across  H$\beta$ and other optical emission lines were also seen in the spectra, with longer time delays observed near the line center and shorter time delays in the line wings.  These initial results indicated that direct modeling would be likely to provide strong constraints on the BLR geometry and kinematics in NGC\,3783, and that we might be able to probe both the low-ionization BLR through the broad H$\beta$ emission line as well as the high-ionization BLR through the \ion{He}{2} $\lambda4686$ broad line.  Here, we present the results of that modeling and a new direct constraint on the black hole mass in NGC\,3783.

\section{Data} 

A detailed description of the photometric and spectroscopic monitoring data are provided by \citet{bentz21a}.  In summary, $V-$band photometric monitoring was carried out with the Las Cumbres Observatory global telescope (LCOGT) network of 1-m telescopes from 12 February to 30 June 2020. Notwithstanding the sudden onset of a global pandemic and the shutdown of several observatories, 209 images were acquired over this period with a median temporal sampling of 0.4\,days.  Spectroscopic monitoring with the robotic FLOYDS spectrograph on the 2-m Faulkes Telescope South was carried out over the same period, with 50 spectra acquired between 27 February and 26 June 2020, with a median temporal sampling of 1.7\,days.

\begin{figure}[t!]
    \epsscale{1.17} 
    \plotone{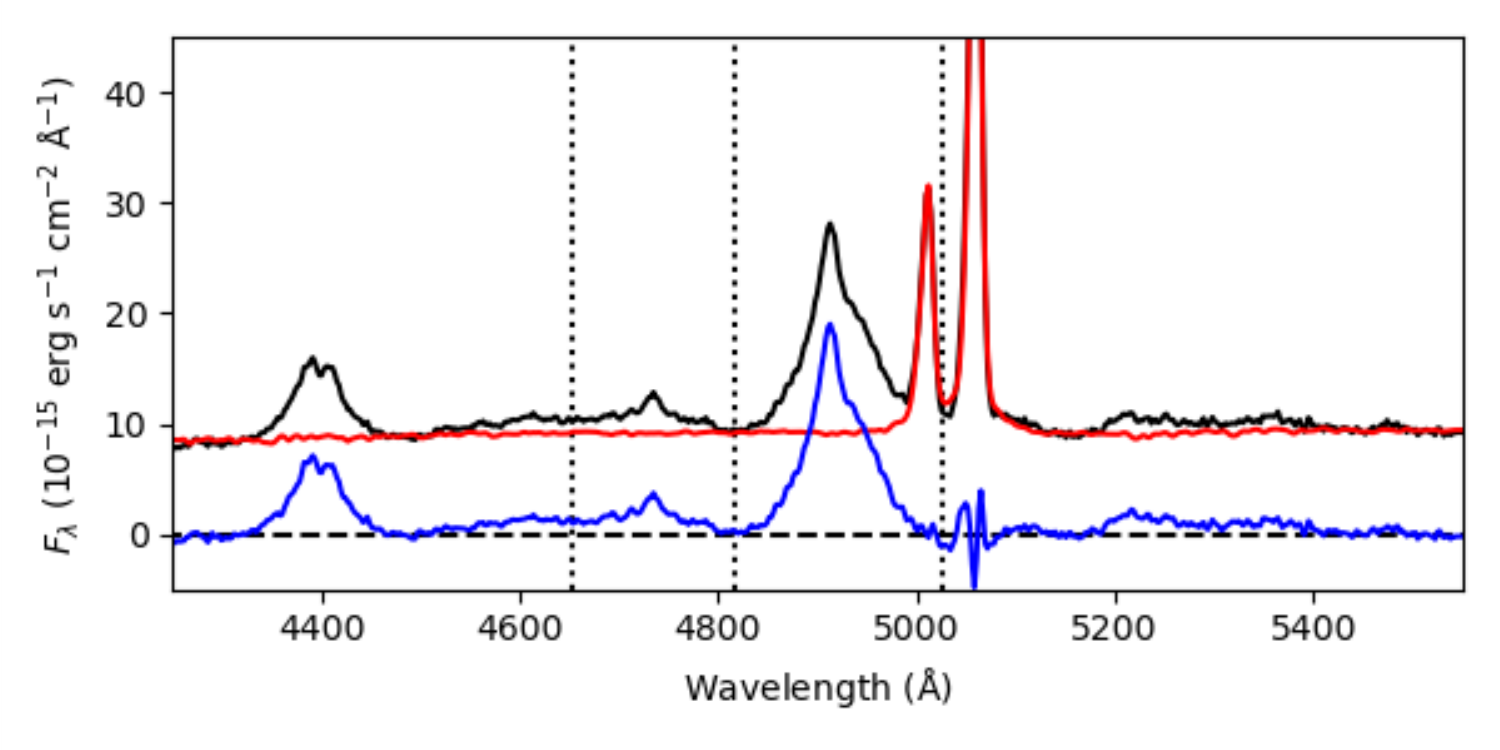}
    \caption{Example spectrum of NGC\,3783 in black with the ULySS fit to the continuum and [\ion{O}{3}] $\lambda\lambda 4959,5007$ doublet overplotted in red, and the continuum and \ion{O}{3}-subtracted spectrum in blue. The vertical dotted lines mark the limits of the regions that were modeled for the H$\beta$  (4816$-$5025\,\AA) and \ion{He}{2} (4653$-$4816\,\AA) emission lines. With the continuum subtracted, low-level \ion{Fe}{2} emission is visible on the blue side of \ion{He}{2} and the red side of the [\ion{O}{3}] doublet, but the analysis of \citet{bentz21a} shows that \ion{Fe}{2} was not variable at a detectable level in these data.}
    \label{fig:specfit}
\end{figure}

The images and spectra were reduced in IRAF\footnote{IRAF is distributed by the National Optical Astronomy Observatories, which are operated by the Association of Universities for Research in Astronomy, Inc., under cooperative agreement with the National Science Foundation.} following standard procedures.  The spectra were intercalibrated using the [\ion{O}{3}] $\lambda\lambda 4959,5007$ emission lines, which are constant in flux on timescales of a few months \citep{peterson13}, thus providing a correction for small wavelength shifts, differences in resolution, and offsets in flux calibration from night to night.   Image subtraction methods \citep{alard98,alard00} were used to isolate the variable AGN point source from the constant host-galaxy emission in the images, providing a well-sampled and well-calibrated light curve of the AGN optical continuum emission.  This was merged with the flux-calibrated continuum light curve measured at $5100\times(1+z)$\,\AA\ in the spectra, with data points taken within 0.25\,days binned together for the final continuum light curve. 

Before modeling the reverberation response, the continuum and [\ion{O}{3}]  emission lines were subtracted from each spectrum to allow the broad emission to be isolated.  This was accomplished by modeling the spectral components in the high signal-to-noise mean spectrum with {\tt ULySS} \citep{koleva09} and then slightly adjusting that model to create a best fit for each individual spectrum before subtracting the desired model components.  The continuum was fit with a powerlaw, representing the AGN continuum contribution, and a host-galaxy component parameterized by the Vazdekis models derived from the MILES library of empirical stellar spectra \citep{vazdekis10}.  Emission lines were fit with multiple Gaussian profiles, with 4 Gaussians needed to match each of the H$\beta$ and [\ion{O}{3}] doublet lines and $1-4$ Gaussians needed to match other emission features in the spectrum.  Once a best fit was achieved for the mean spectrum, the individual spectra were then fit one at a time, with the host-galaxy component held fixed to the best-fit template but allowed to vary in flux contribution, and with the power law and the emission-line components allowed to vary but with initial values matching their best-fit values.  Once a best fit was found, the host-galaxy and power law continua and the [\ion{O}{3}] components were then subtracted from each spectrum.  Figure~\ref{fig:specfit} shows an example spectrum from a single night of observations in black, with the best-fit continuum and [\ion{O}{3}] emission in red, and the spectrum after subtraction of those components in blue.

The H$\beta$ region was then isolated for modeling between observed wavelengths $4816-5025$\,\AA\  with the narrow emission line peak at 4910\,\AA, while the \ion{He}{2} region was isolated between $4653-4816$\,\AA\ with the narrow emission line peak observed at 4735\,\AA.   Throughout the campaign, the rest-frame equivalent width of broad H$\beta$ relative to the starlight-corrected AGN continuum has a mean value of 139.9\,\AA\ with a median  of 130.5\,\AA\ and a standard deviation of 22.4\,\AA.   For \ion{He}{2}, the mean rest-frame equivalent width is 15.8\,\AA\ with a median of 15.1\,\AA\ and a standard deviation of 5.4\,\AA.
While the blue spectra also cover the H$\gamma$ and H$\delta$ broad emission lines, and the red spectra cover the H$\alpha$ emission line, \citet{bentz21a} described the difficulties in accurately calibrating the red spectra and the short wavelength end of the blue spectra.  The integrated light curves for these emission lines clearly demonstrate significant excess noise, so we do not attempt to model them here.

\section{BLR Models}

Modeling of the BLR for H$\beta$ and for \ion{He}{2} was carried out with {\tt CARAMEL}, a phenomenological modeling code that is described in detail by \citet{pancoast14a}. {\tt CARAMEL} is capable of constraining both the geometry and kinematics of the BLR using the reverberation response across the profile of a broad emission line throughout a monitoring campaign.  Here, we summarize the main components of the model.

{\tt CARAMEL} represents the BLR as a large collection of massless point particles that are distributed in position and velocity space,  surrounding a massive black hole whose gravity dominates the region.  Each point particle processes incident continuum flux instantaneously, and the observed time delay profile of the BLR depends on the spatial distribution of point particles while the broad line wavelength profile depends on the velocity distribution of point particles.

The spatial distribution of particles is parametrized with angular and radial distributions. The radial positions of particles are drawn from a gamma distribution
\begin{equation}
    p(x|\alpha,\theta) \propto x^{\alpha - 1}\exp{\left( - \frac{x}{\theta} \right)}
\end{equation}
\noindent that provides the flexibility to represent a Gaussian  ($\alpha>1$), an exponential ($\alpha=1$), or a cuspier profile ($0<\alpha<1$).  The gamma distribution of particles is shifted away from the location of the black hole by the Schwarzschild radius, $R_s = 2GM/c^2$, plus a minimum radius $r_{\rm min}$. To assist with interpretation of the modeling results, a change of variables is performed so that parametrization is given in terms of ($\mu$, $\beta$, $F$):
\begin{equation}
    \mu = r_{\rm min} + \alpha \theta,
\end{equation}
\begin{equation}
    \beta = \frac{1}{\sqrt{\alpha}},
\end{equation}
\begin{equation}
    F = \frac{r_{\rm min}}{r_{\rm min} + \alpha \theta},
\end{equation}
\noindent where $\mu$ is the mean radius, $\beta$ is the shape parameter, and $F$ is $r_{\rm min}$ in units of $\mu$.  The standard deviation of the shifted gamma profile is given by $\sigma_r = \mu \beta (1-F)$, and the BLR is truncated at an outer radius of $r_{\rm out} = c \Delta t_{\rm data}/2$, where $\Delta t_{\rm data}$ is the time difference between the first point in the modeled continuum light curve and the first point in the emission-line light curve.  This truncation assumes that the total length of the monitoring campaign is sufficient to track reverberation signals throughout the entire BLR.

The angular distribution of the particles is then arranged in a disk with a thickness that is set by an opening angle $\theta_o$, where $\theta_o=0\degr$ is a thin disk and $\theta_o=90\degr$ is a sphere.  The inclination of the disk to the observer's line of sight is set by $\theta_i$, where $\theta_i=0\degr$ is viewed face on and $\theta_i=90\degr$ is viewed edge on.  The strength of line emission from different depths within the disk is parametrized by the distribution of particles as a function of depth.  For a single particle, the angle of displacement from the disk midplane is given by
\begin{equation}
    \theta_{d,N} = \arccos (\cos \theta_o + (1-\cos \theta_o)\times U^{\gamma})
\end{equation}
\noindent where $U$ is a random number drawn uniformly between 0 and 1.  The value of $\gamma$ ranges from 1, where particles are distributed uniformly throughout the thickness of the disk, to 5, where particles are clustered at the disk face and therefore emission is preferentially from the outer skin of the BLR.  An additional asymmetry parameter, $\xi$ allows for the possibility of obscuration along the midplane of the disk, where $\xi \rightarrow 0$ causes the entire back half of the disk to be obscured and $\xi = 1$ has no midplane obscuration.  The final asymmetry parameter $\kappa$ is related to the weight of a particle 
\begin{equation}
    W(\phi) = \frac{1}{2} + \kappa \cos \phi
\end{equation}
where $W$ is the fraction of continuum flux that is reradiated back towards the observer as line flux  and $\phi$ is the angle between the observer's line of sight to the source and the particle's line of sight to the source. The value of $\kappa$ ranges from $-0.5$, where particles preferentially emit back towards the ionizing source, to $0.5$, where particles preferentially radiate away from the ionizing source.  In the case of $\kappa=-0.5$, an observer would see preferential emission from the far side of the disk, while preferential emission from the near side would be observed in the case of $\kappa=0.5$.

\begin{figure*}
    \epsscale{1.17} 
    \plotone{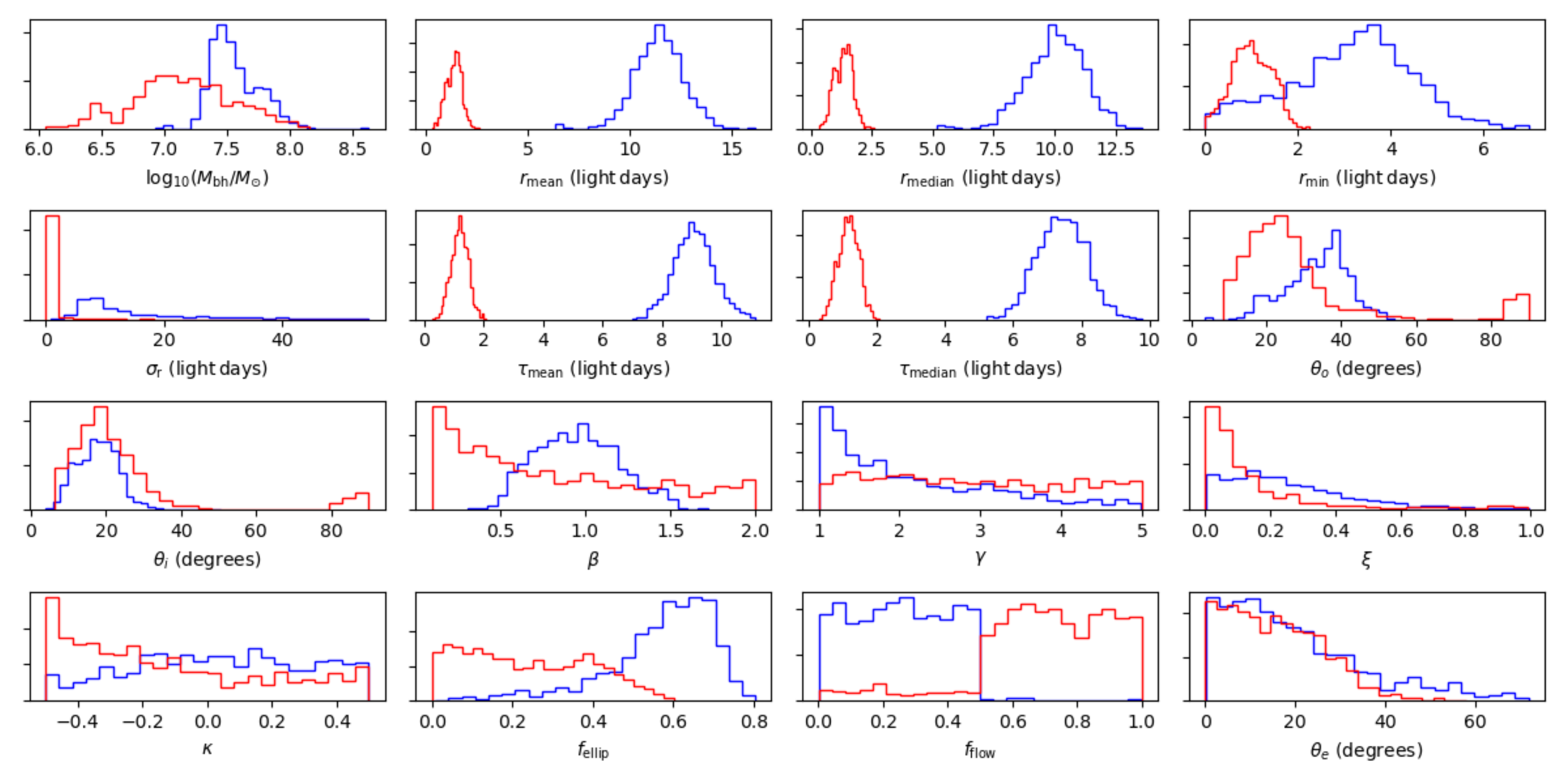}
    \caption{
    Histograms displaying the posterior distributions of the BLR model parameters for H$\beta$ (blue) and \ion{He}{2} (red).
    }
    \label{fig:modelpars}
\end{figure*}

\begin{deluxetable*}{LlCC}
\renewcommand{\arraystretch}{1.1}
\tablecolumns{4}
\tablecaption{Broad-line region model parameter values}
\tablehead{
\colhead{Parameter} & 
\colhead{Brief Description} & 
\colhead{H$\beta$} &
\colhead{\ion{He}{2}}
}
\startdata
\log_{10} (M/M_{\odot})         & Black hole mass                       & 7.51^{+0.26}_{-0.13}      &  7.13^{+0.43}_{-0.37} \\
r_{\rm mean}\ \rm(light~days)   & Mean radius of line emission          & 11.41^{+1.13}_{-1.17}     &  1.40^{+0.31}_{-0.42} \\ 
r_{\rm median}\ \rm(light~days)  & Median radius of line emission        & 10.07^{+1.10}_{-1.21}     &  1.33^{+0.34}_{-0.42} \\
r_{\rm min}\ \rm(light~days)     & Minimum radius of line emission       & 3.25^{+1.13}_{-1.54}      &  1.00^{+0.46}_{-0.42} \\
\sigma_{r}\ \rm(light~days)      & Radial extent of line emission        & 10.47^{+15.44}_{-3.82}    &  0.17^{+0.34}_{-0.13} \\
\tau_{\rm mean}\ \rm(days)       & Mean time delay                       & 9.05^{+0.68}_{-0.64}      &  1.19^{+0.28}_{-0.30} \\
\tau_{\rm median}\ \rm(days)     & Median time delay                     & 7.42^{+0.70}_{-0.74}      &  1.16^{+0.29}_{-0.32} \\
\theta_{o}\ \rm(degrees)        & Opening angle                         & 34.7^{+6.2}_{-9.9}        &  23.5^{+11.8}_{-8.0} \\
\theta_{i}\ \rm(degrees)        & Inclination angle                     & 17.9^{+5.3}_{-6.1}        &  19.1^{+10.3}_{-7.0}  \\
\beta                           & Shape parameter of radial distribution & 0.95^{+0.25}_{-0.25}     &  0.67^{+0.83}_{-0.45} \\
\gamma                          & Disk face concentration parameter     & 1.84^{+1.48}_{-0.67}      &  2.77^{+1.55}_{-1.23} \\
\xi                             & Transparency of the mid-plane         & 0.23^{+0.24}_{-0.15}      &  0.08^{+0.23}_{-0.06} \\
\kappa                          & Cosine illumination function parameter & 0.04^{+0.31}_{-0.30}     &  -0.20^{+0.45}_{-0.24} \\
f_{\rm ellip}                   & Fraction of elliptical orbits         & 0.60^{+0.09}_{-0.15}      &  0.22^{+0.19}_{-0.16} \\
f_{\rm flow}                    & Inflow vs.\ outflow                   & 0.26^{+0.17}_{-0.18}      &  0.72^{+0.19}_{-0.17} \\
\theta_{e}\ \rm(degrees)         & Ellipse angle                         & 16.1^{+18.6}_{-11.0}      &  14.6^{+11.8}_{-10.3} \\
\sigma_{\rm turb}               & Turbulence                            & 0.024^{+0.050}_{-0.021}   &  0.013^{+0.044}_{-0.011} \\
r_{\rm out}\ \rm(light~days)     & Outer radius of line emission (fixed parameter) & 42              & 42 \\
T                               & Temperature or likelihood softening   & 125                       & 145
\label{tab:modelpars}
\enddata

\tablecomments{Tabulated values are the median and 68\% confidence intervals.}
\end{deluxetable*}

The velocity distribution of particles includes radial and tangential distributions.  A fraction of the particles, $f_{\rm ellip}$, have near-circular orbits within the Keplerian potential of the central black hole with mass \mbh.  The remaining particles ($1-f_{\rm ellip}$) are either inflowing ($f_{\rm flow}<0.5$) or outflowing ($f_{\rm flow}>0.5$). Whether these orbits are generally bound or unbound is determined by the parameter $\theta_e$.  For a plane defined by the possible values of the radial and tangential velocities, $\theta_e$  describes the angle of the velocity components away from the escape velocity and towards the circular velocity.  If $\theta_e =0$\,degrees then the  orbits are drawn from a Gaussian distribution centered on the escape velocity. As $\theta_e \rightarrow 90\degr$, the inflowing or outflowing orbits approach the parameter space occupied by near-circular orbits.  Thus high values of $\theta_e$ indicate inflowing or outflowing orbits that are very nearly circular, $\theta_e\approx45\degr$ indicates that most of the inflowing or outflowing orbits are highly eccentric but still bound, and low values of $\theta_e$ indicate that most particles are near the escape velocity and  unbound.

A contribution from macroturbulence is included in the line-of-sight component of the velocity vector for each point particle as 
\begin{equation}
    v_{\rm turb} = \mathcal{N} (0,\sigma_{\rm turb})|v_{\rm circ}|,
\end{equation}
\noindent where $v_{\rm circ}$ is the circular velocity and $\mathcal{N}(0,\sigma_{\rm turb})$ is a normal distribution centered on 0 and with standard deviation $\sigma_{\rm turb}$.

With the spatial and velocity distributions of the particles parametrized, the emission-line profile can then be calculated for each continuum flux measurement, assuming that the continuum flux tracks the ionizing flux from a central point source. A nonvariable narrow emission-line component is included in the modeled emission-line profiles, as is a smoothing parameter to account for the small differences in spectral resolution that arise from variable seeing conditions throughout the monitoring campaign.

To explore the full range of possible time delays arising from the BLR geometry and to properly compare the modeled emission line profiles with the measured profiles, the continuum light curve must be interpolated.  {\tt CARAMEL} uses Gaussian processes to both interpolate between continuum flux measurements and to extrapolate the continuum light curve beyond the beginning and end of the monitoring campaign to extend the range of time delays that may be probed.  The uncertainties on the Gaussian process model parameters are included in the determination of the BLR model parameters, thus capturing the effect of the uncertainties that arise from interpolating and extrapolating the continuum data.

For each model realization, we include 2000 individual point particles to represent the BLR.  The continuum light curve is interpolated and model emission-line profiles are calculated for each epoch at which an emission-line measurement was acquired.  A Gaussian likelihood function compares the modeled spectra against the measured spectra and adjusts the model parameters accordingly. {\tt CARAMEL} utilizes a diffusive nested sampling code, with the latest version employing {\tt DNEST4} \citep{brewer18}, to efficiently explore the model parameter space.  {\tt DNEST4} allows for the use of a likelihood softening parameter, or statistical temperature $T$, which has the effect of increasing the measurement uncertainties.  This parameter can account for underestimated measurement uncertainties or for the inability of the simplified model to capture all of the real details in the measurements.  The value of $T$ is determined in the post analysis by examining the distributions of the model parameters and choosing the largest value of $T$ for which the distributions remain smooth and generally unimodal.

Finally, to check that convergence had been reached, we compared the constrained values of the model parameters from the first half of the model runs to the second half of the model runs, with the total number of model runs being 10,000.  There was no discernible difference between the parameters constrained during the first half or second half of the model runs for either H$\beta$ or \ion{He}{2}.

\section{Results}

\begin{figure}
    \epsscale{1.17} 
    \plotone{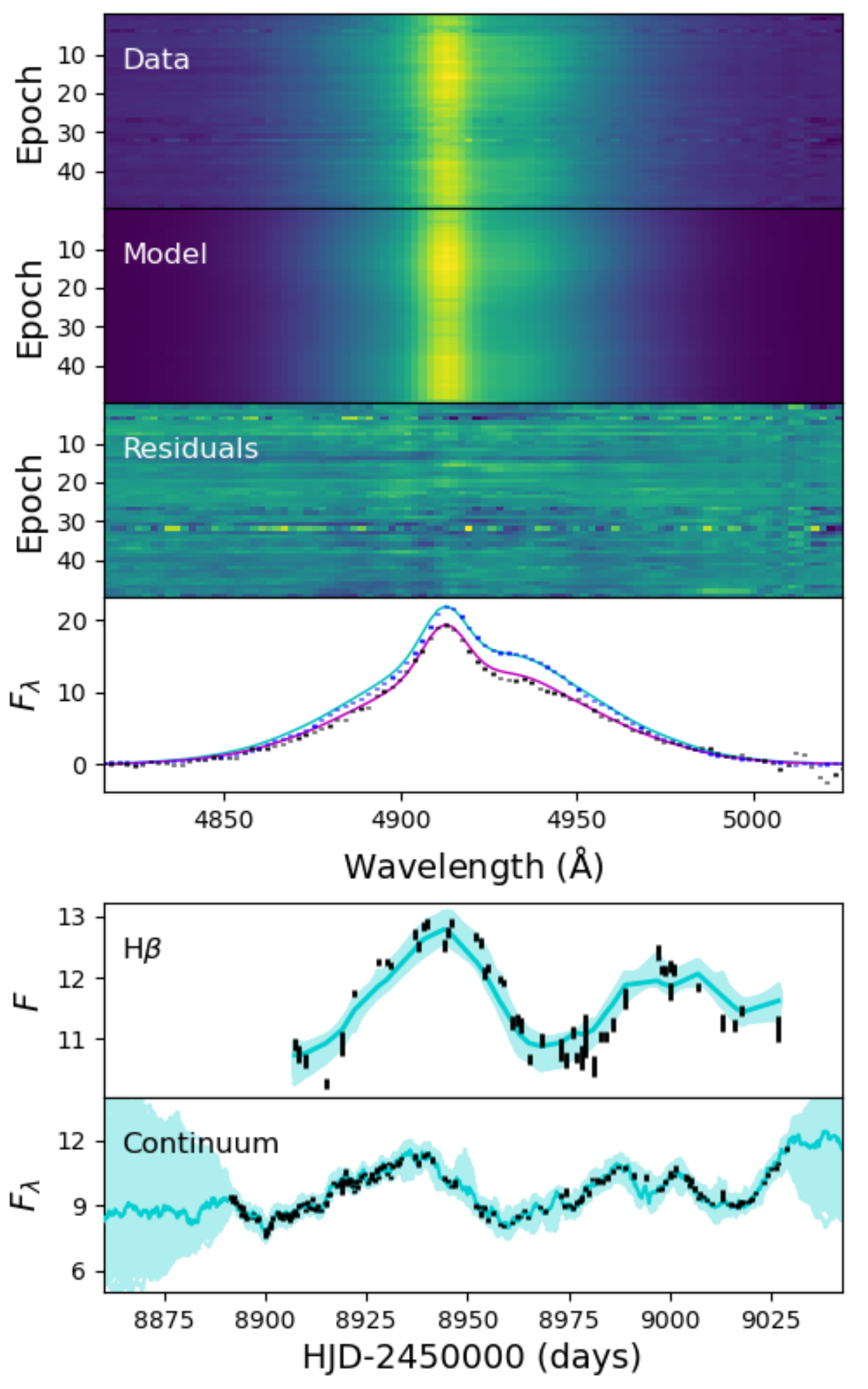}
    \caption{
    The top three panels display the data, one possible model, and residuals (data$-$model) for the H$\beta$ spectra, with epochs 1 and 13 and their model fits displayed immediately below to exemplify a low flux spectrum (magenta curve) and a high flux spectrum (cyan curve). The bottom two panels display the continuum and integrated H$\beta$ light curves as data points with model fits overlaid.  The full ranges of the models are displayed in light turquoise with the example model corresponding to the top four panels overlaid in dark turquoise.  Flux densities ($F_{\lambda}$) are in units of $10^{-15}$\,erg\,s$^{-1}$\,cm$^{-2}$\,\AA$^{-1}$ while integrated flux ($F$) is in units of $10^{-13}$\,erg\,s$^{-1}$\,cm$^{-2}$. Across the six panels, it is evident that most of the gross characteristics of the data are captured by the models, although some of the finer details are not. Furthermore, the continuum model is less well constrained during time periods with multi-day gaps between the measurements.  Unfortunately, these gaps resulted from the shutdown of numerous observatories in response to the global coronavirus pandemic and could not be avoided.
    }
    \label{fig:hbfits}
\end{figure}

Modeling of the H$\beta$ emission line in NGC\,3783 provides constraints on the low ionization BLR, while modeling of the \ion{He}{2} emission line constrains the high ionization BLR.  Figure~\ref{fig:modelpars} compares the posterior probability distribution functions for all the parameters of the BLR models for both H$\beta$ and \ion{He}{2}, while the median and 68\% confidence intervals for each parameter are listed in Table~\ref{tab:modelpars}.  We describe the resultant set of models for each emission line below.

\subsection{H$\beta$} 

The models for H$\beta$ require a likelihood softening of $T=125$, which amounts to increasing the uncertainties on the data by a factor of $\sqrt{T} = 11.2$.  Figure~\ref{fig:hbfits} displays the continuum and integrated H$\beta$ emission-line light curves and the observed H$\beta$ line profiles along with model fits to all of these.  In general, the emission line profiles are well-fit by the modeled profiles as are the gross flux variations of the integrated emission-line light curve, although some of the finer details of the data are not captured by the models.  The small disagreements between the data and the models could be the result of uncertainties that are still underestimated for some data points, or they could signal that the models are too simplistic and do not have the full flexibility needed to match all of the real variations, or both.

\begin{figure}
    \epsscale{1.17} 
    \plotone{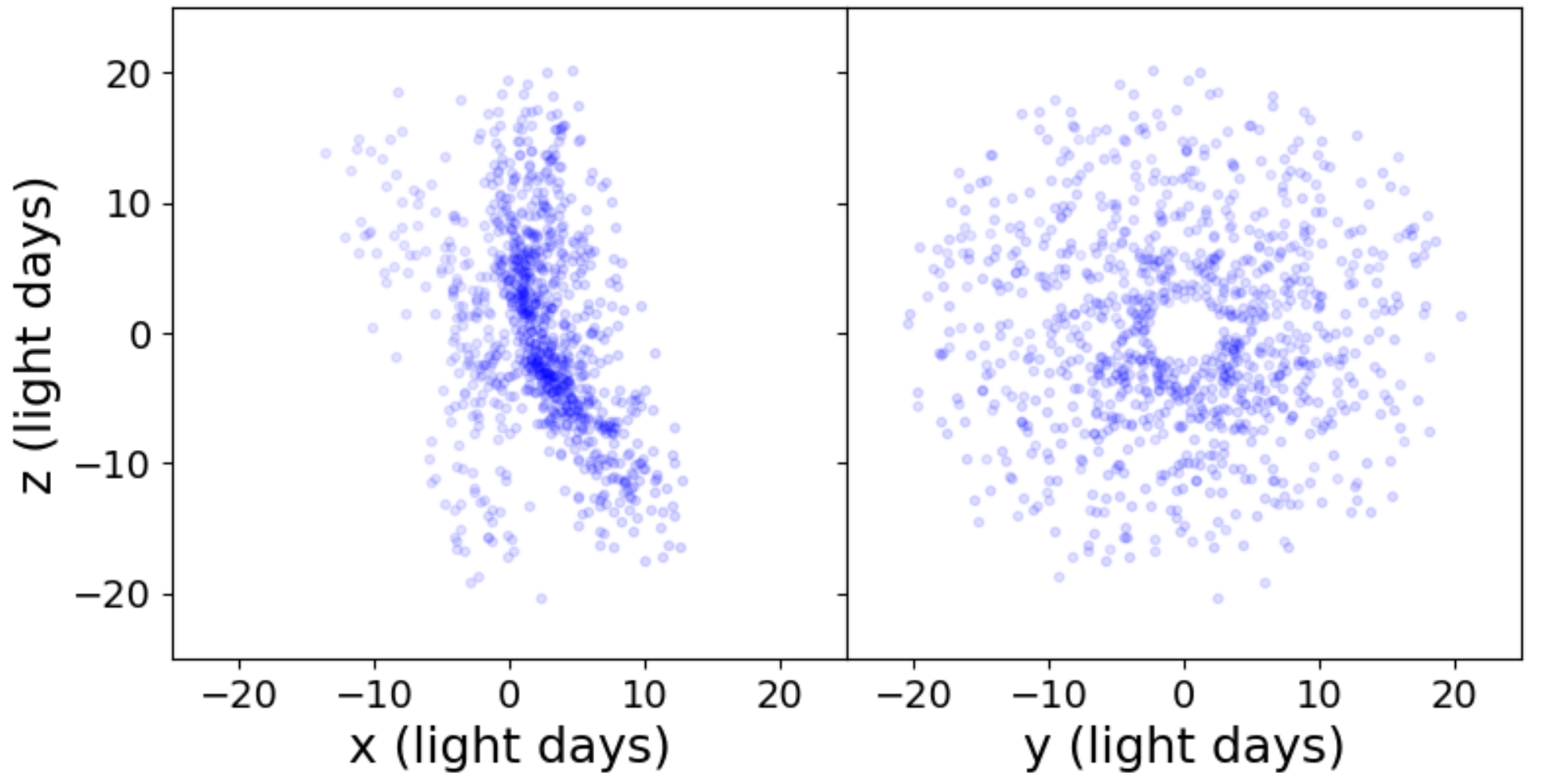}
    \caption{
    Representative geometric model for the H$\beta$ response in the broad line region of NGC\,3783, drawn from the posterior probability distribution.  The left panel is oriented edge on, with an Earth-based observer on the +x axis, while the right panel shows the Earth-based observer's view.  The transparency of each point represents the relative response of the gas to continuum fluctuations at each position, with more opaque points responsible for a stronger response. This effect is most easily viewed in the right panel, where there is less overlap between points.
    }
    \label{fig:hbclouds}
\end{figure}

The geometry of the H$\beta$-emitting BLR is found to be a relatively face-on thick disk with an opening angle of $\theta_o=34.7^{+6.2}_{-9.9}$\,degrees and an inclination to our line of sight of $\theta_i=17.9^{+5.3}_{-6.1}$\,degrees.  The disk has an inner minimum radius of $r_{\rm min}=3.25^{+1.13}_{-1.54}$\,light~days with a median radius of $r_{\rm median}=10.07^{+1.10}_{-1.21}$\,light~days and a width of $\sigma_r=10.47^{+15.44}_{-3.82}$\,light~days.  The disk emission is distributed radially in a near-exponential profile ($\beta=0.95^{+0.25}_{-0.25}$), and is distributed throughout the thickness of the disk with a slight preference for stronger emission near the face of the disk ($\gamma=1.84^{+1.48}_{-0.67}$) and strong but not total obscuration along the midplane ($\xi=0.23^{+0.24}_{-0.15}$).  The line emission direction is rather unconstrained, with the median value centered around isotropic emission but having large uncertainties that do not discriminate between a preference for radiation towards or away from the central source ($\kappa=0.04^{+0.31}_{-0.30}$).  Figure~\ref{fig:hbclouds} displays a representative geometric model for the H$\beta$ response in the BLR of NGC\,3783, drawn from the posterior probability distribution.  

\begin{figure}
    \epsscale{1.17} 
    \plotone{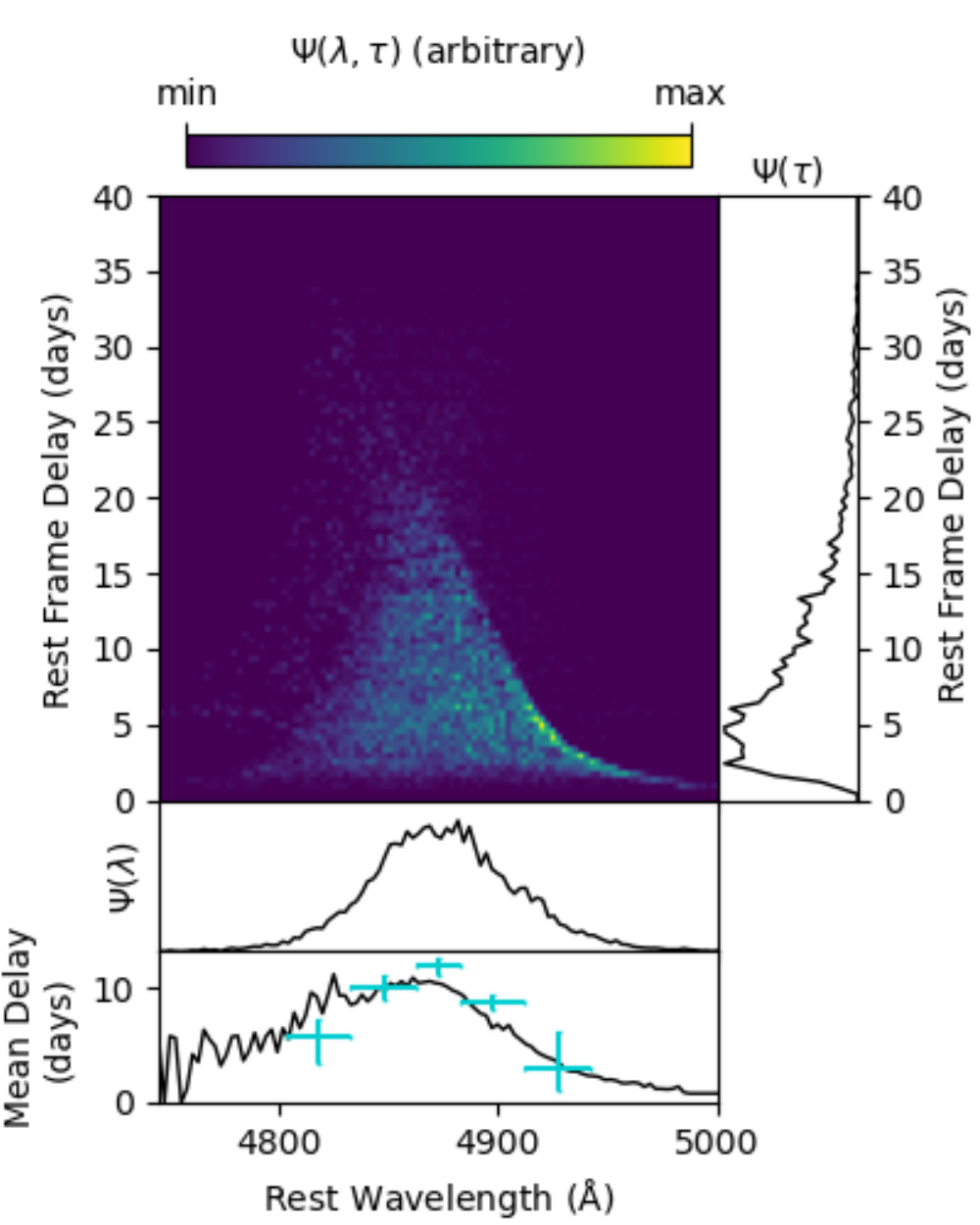}
    \caption{
    Transfer function $\Psi(\lambda,\tau)$ for the example H$\beta$ model displayed in Figure~\ref{fig:hbclouds}.  Integrating the transfer function over wavelength gives the one-dimensional lag profile $\Psi(\tau)$, which is shown on the right.  Integrating the transfer function over time delay gives $\Psi(\lambda)$, or the variable emission-line profile, which is shown immediately under the transfer function.  The bottom panel displays the average lag as a function of wavelength across the emission line, with the turquoise crosses showing the average time delay for 5 velocity bins across the H$\beta$ profile from Figure~6 of \citet{bentz21a}.
    }
    \label{fig:hbtransfer}
\end{figure}

The associated mean and median time delays for H$\beta$ are found to be $\tau_{\rm mean}=9.05^{+0.68}_{-0.64}$\,days and $\tau_{\rm median}=7.42^{+0.70}_{-0.74}$\,days, which agree well with the average H$\beta$ time delay reported by \citet{bentz21a} of $\tau_{\rm cent}=9.60^{+0.65}_{-0.72}$\,days.  Figure~\ref{fig:hbtransfer} displays the  transfer function, $\Psi(\lambda,\tau)$, for a representative model.  Also referred to as the velocity-delay map, the transfer function displays the range of H$\beta$ responsivities as a function of time delay and velocity (or wavelength) across the broad emission line profile.  The shape of the transfer function generally agrees with the cross-correlation time delays computed for different velocity bins of the H$\beta$ profile by \citet{bentz21a}, displayed here as the turquoise crosses in the bottom panel of Figure~\ref{fig:hbtransfer}. 

The black hole mass is constrained to be $\log_{10} (M_{\rm BH}/M_{\odot})=7.51^{+0.26}_{-0.13}$.  Roughly 60\% of the particle orbits are near circular ($f_{\rm ellip}=0.60^{+0.09}_{-0.15})$, with the other 40\% strongly preferring inflow ($f_{\rm flow}<0.5$).  With a low value of $\theta_e=16.1^{+18.6}_{-11.0}$\,degrees, most of these are truly inflowing orbits rather than highly elliptical bound orbits.  There is also a small but non-zero contribution to the kinematics from turbulence ($\sigma_{\rm turb}=0.024^{+0.050}_{-0.021}$).

\subsection{\ion{He}{2}}

\begin{figure}
    \epsscale{1.17} 
    \plotone{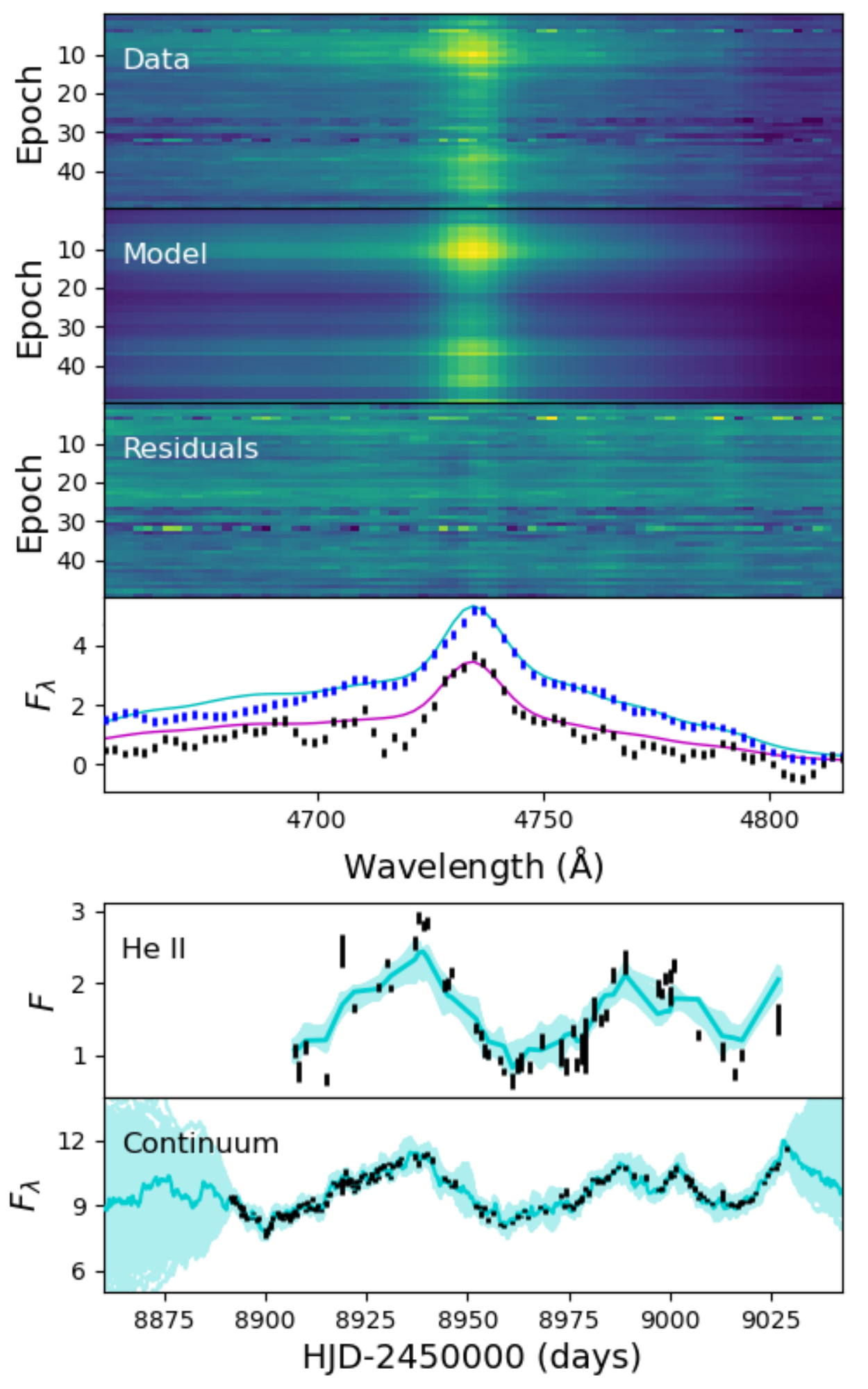}
    \caption{
    Same as Figure~\ref{fig:hbfits}, but for \ion{He}{2}. 
    }
    \label{fig:he2fits}
\end{figure}

The models for \ion{He}{2} require a likelihood softening of $T=145$, which amounts to increasing the uncertainties on the data by a factor of $\sqrt{T} = 12.0$.  Figure~\ref{fig:he2fits} displays the continuum and integrated \ion{He}{2} emission-line light curves and the observed \ion{He}{2} line profiles along with model fits to all of these.  In general, the modeled emission line profiles fit the main features of the observations, however the lower integrated flux and larger uncertainties compared to H$\beta$ do lead to somewhat less agreement between the observations and the models. The gross flux variations of the integrated emission-line light curve also seem to be mostly captured by the models.

\begin{figure}
    \epsscale{1.17} 
    \plotone{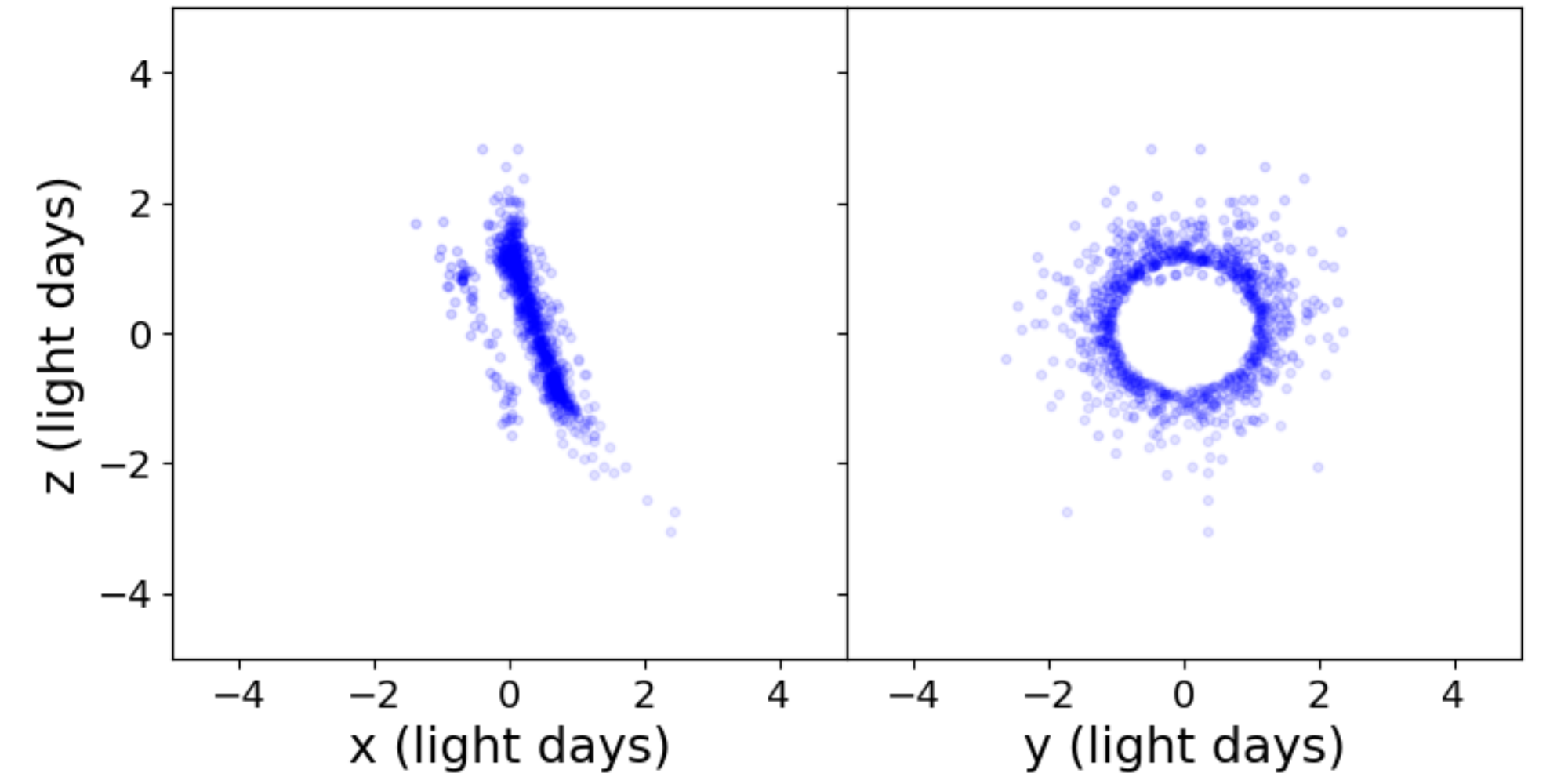}
    \caption{
    Same as Figure~\ref{fig:hbclouds}, but for \ion{He}{2}. 
    }
    \label{fig:he2clouds}
\end{figure}

The geometry of the \ion{He}{2}-emitting BLR is again found to be a relatively face-on thick disk  with an opening angle of $\theta_o=23.5^{+11.8}_{-8.0}$\,degrees and an inclination to our line of sight of $\theta_i=19.1^{+10.3}_{-7.0}$\,degrees.  The disk has an inner minimum radius of $r_{\rm min}=1.00^{+0.46}_{-0.42}$\,light~days with a median radius of $r_{\rm median}=1.33^{+0.34}_{-0.42}$\,light~days and a width of $\sigma_r=0.17^{+0.34}_{-0.13}$\,light~days.  The disk emission is distributed radially in a Gaussian profile ($\beta=0.67^{+0.83}_{-0.45}$), although the constraints on this parameter are quite weak.  The distribution of emission throughout the thickness of the disk is also not well constrained ($\gamma=2.77^{+1.55}_{-1.23}$), but there is a preference for strong obscuration along the midplane ($\xi=0.08^{+0.23}_{-0.06}$).  The line emission slightly prefers radiation back towards the central source ($\kappa=-0.20^{+0.45}_{-0.24}$). 
Figure~\ref{fig:he2clouds} displays a representative model for the \ion{He}{2} response in the BLR of NGC\,3783, drawn from the posterior probability distribution.  As expected, it is significantly more compact than H$\beta$.   

\begin{figure}
    \epsscale{1.17} 
    \plotone{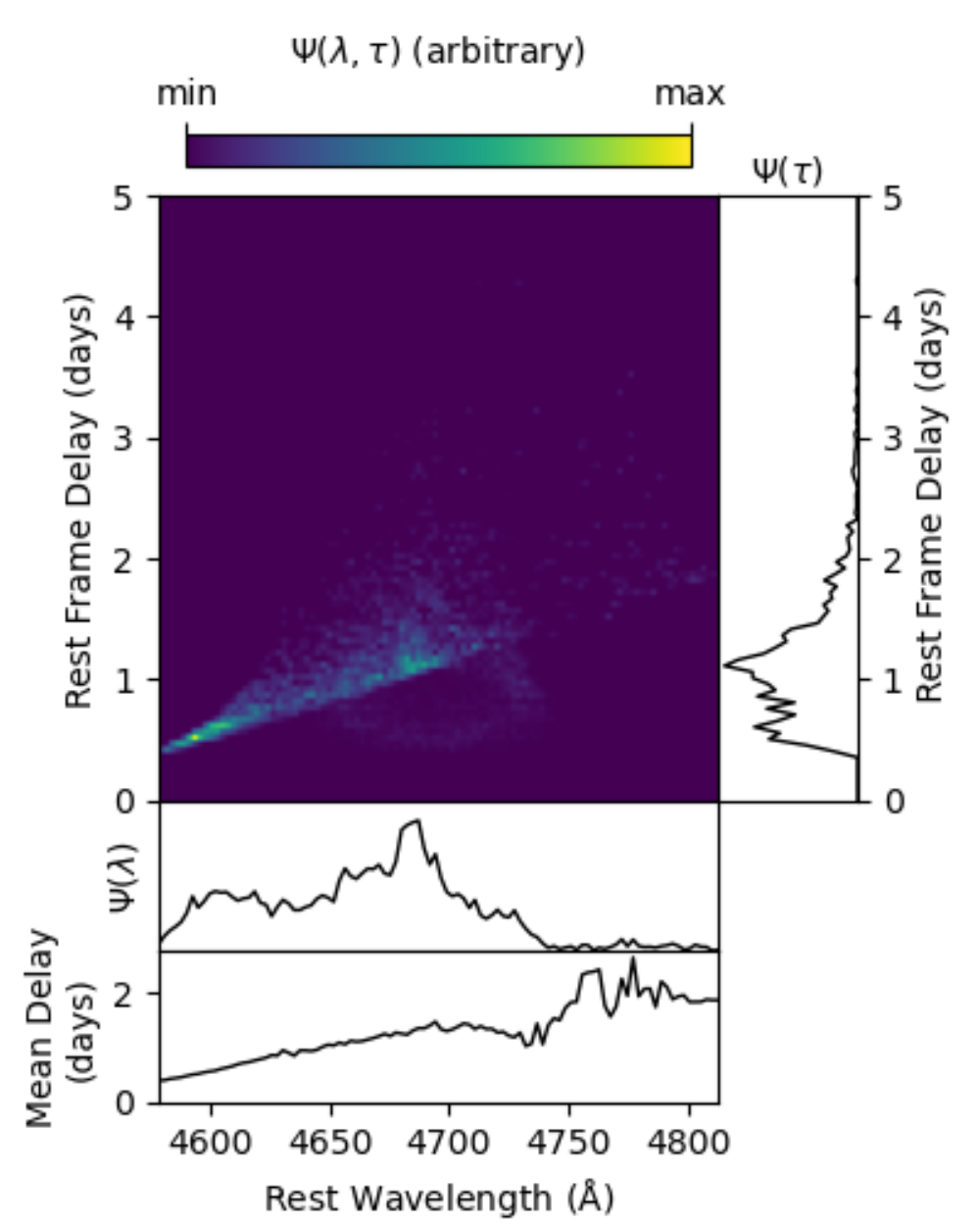}
    \caption{
    Same as Figure~\ref{fig:hbtransfer}, but for \ion{He}{2}. 
    }
    \label{fig:he2transfer}
\end{figure}

The associated mean and median time delays for \ion{He}{2} are found to be $\tau_{\rm mean}=1.19^{+0.28}_{-0.30}$\,days and $\tau_{\rm median}=1.16^{+0.29}_{-0.32}$\,days, which are a bit more compact but agree within the uncertainties with the average \ion{He}{2} time delay reported by \citet{bentz21a} of $\tau_{\rm cent}=1.95^{+1.02}_{-0.98}$\,days.  Figure~\ref{fig:he2transfer} displays the transfer function for a representative model.  The shape is much more asymmetric than was found for H$\beta$, with a heavier response in the blue wing and very little response in the red wing.

The black hole mass is constrained to be $\log_{10} (M_{\rm BH}/M_{\odot})=7.13^{+0.43}_{-0.37}$.  Only $1/5$ of the particle orbits are near circular ($f_{\rm ellip}=0.22^{+0.19}_{-0.16}$), while the rest of the orbits strongly preferring outflow ($f_{\rm flow}>0.5$).  With a low value of $\theta_e=14.6^{+11.8}_{-10.3}$\,degrees, most of these are truly outflowing orbits rather than highly elliptical bound orbits.  Finally, there is again a small but non-zero contribution to the kinematics from turbulence ($\sigma_{\rm turb}=0.013^{+0.044}_{-0.011}$).

\begin{figure}
    \epsscale{1.17} 
    \plotone{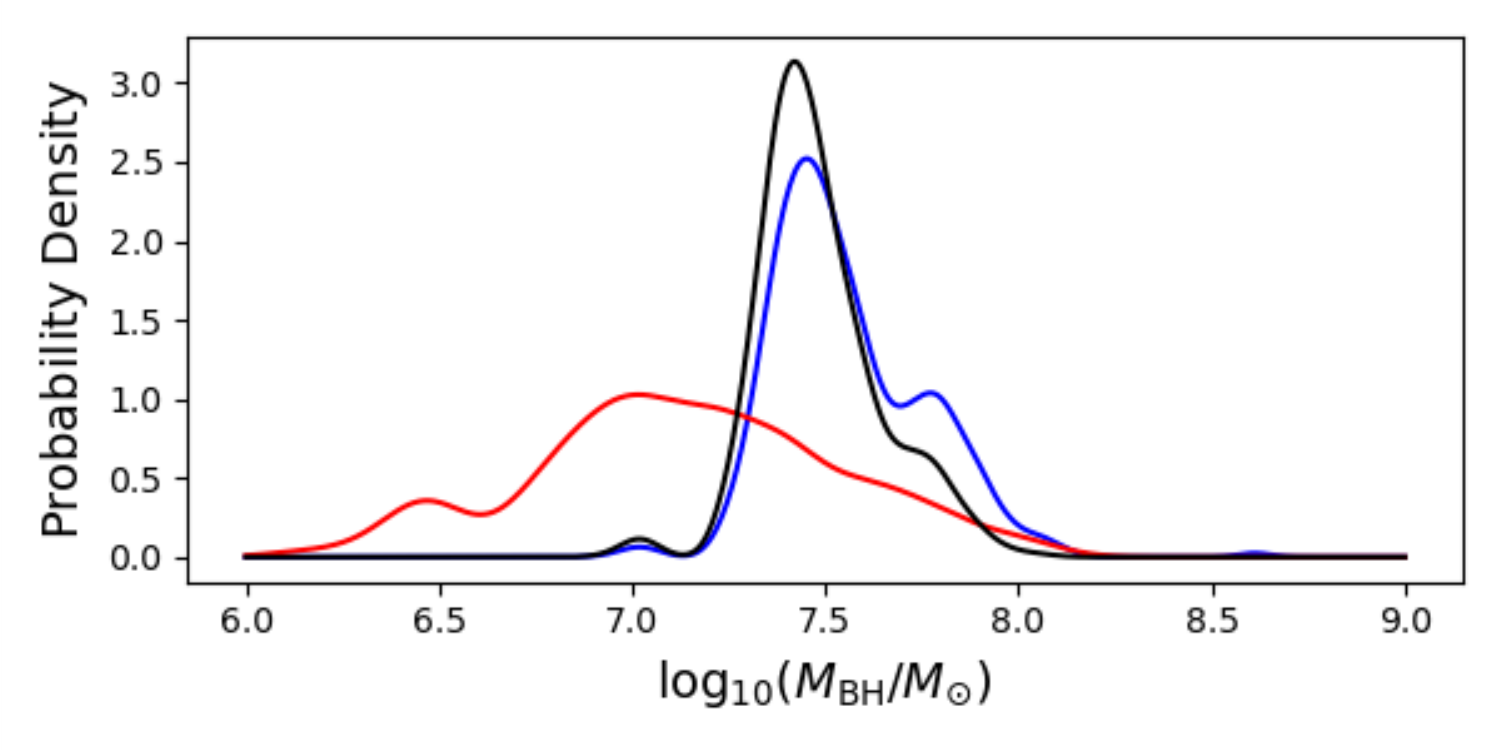}
    \caption{
    Constraints on the black hole mass in NGC\,3783 from H$\beta$ (blue), \ion{He}{2} (red), and the joint inference using results from both emission lines (black).
    }
    \label{fig:mbh}
\end{figure}

\begin{figure*}
    \epsscale{1.17} 
    \plotone{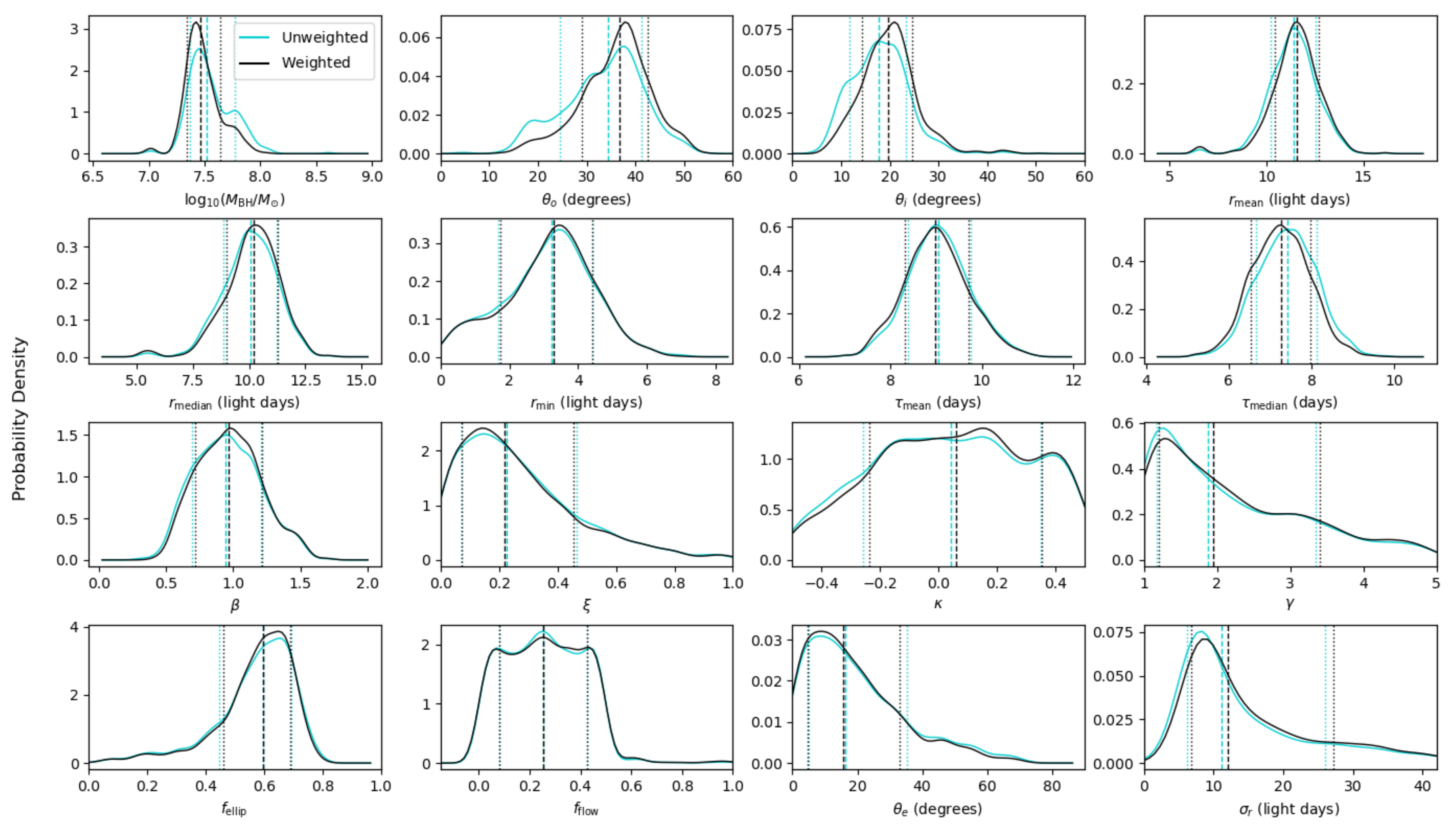}
    \caption{
    Posterior distributions of the BLR model parameters for H$\beta$ before (turquoise, ``unweighted'') and after (black, ``weighted'') selecting only those models that agree with the joint constraint on \mbh. The unweighted distributions in turquoise are the same as the results for H$\beta$ in Figure~\ref{fig:modelpars}, but are effectively smoothed with a Gaussian kernel for easier comparison with the weighted constraints.  The vertical dotted lines mark the median values, while the dashed vertical lines mark the 68\% confidence intervals.  The parameters are generally unchanged when models that agree with the joint constraint on \mbh\ are preferred.
    }
    \label{fig:hb_mbh_weight}
\end{figure*}

\begin{figure*}
    \epsscale{1.17} 
    \plotone{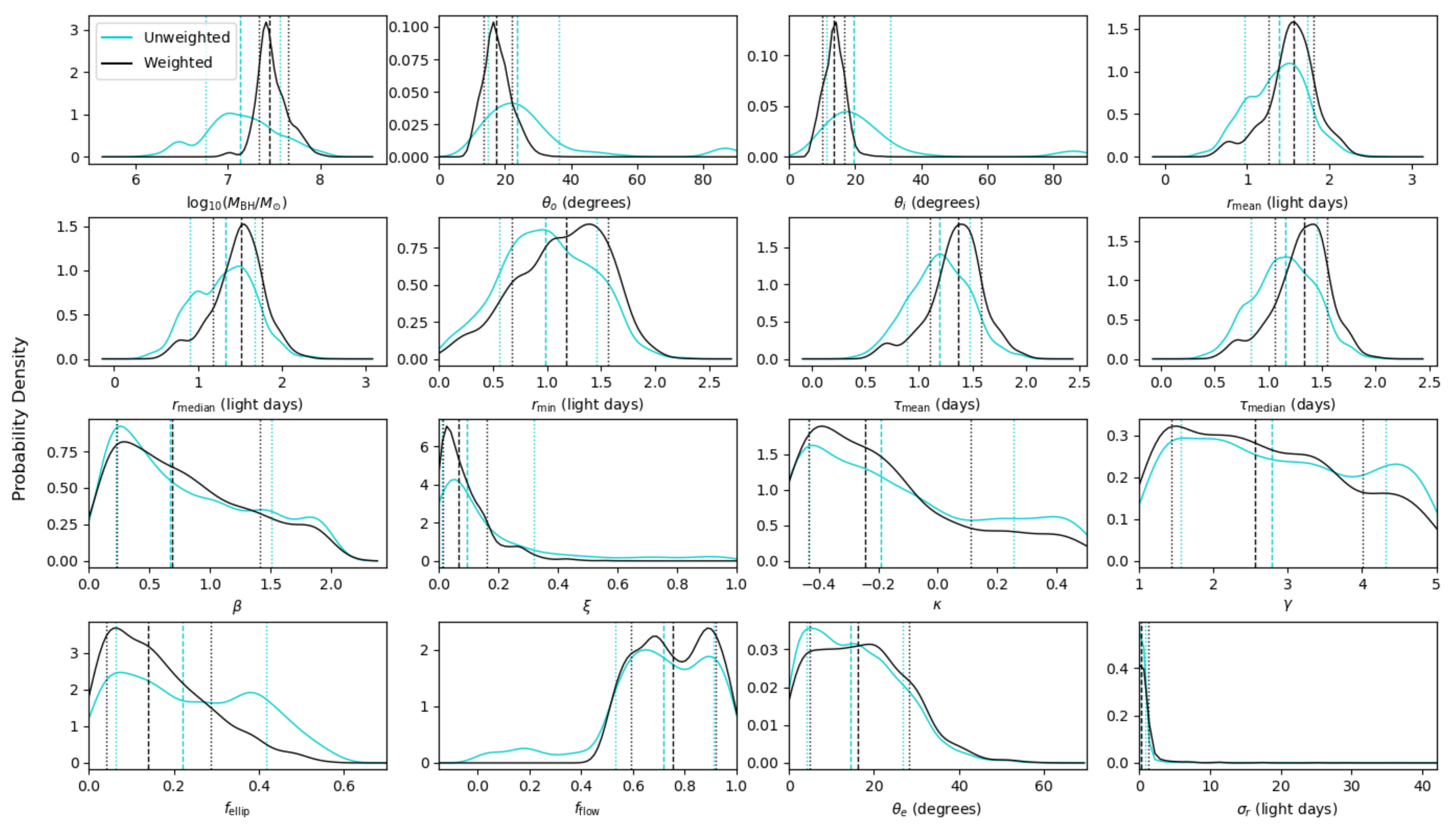}
    \caption{
    Same as Figure~\ref{fig:hb_mbh_weight} but for \ion{He}{2}. The most significant changes are for the BLR radius, which shifts to larger values, and the fraction of near-circular orbits, which decreases to smaller values.
    }
    \label{fig:he2_mbh_weight}
\end{figure*}

\section{Discussion}

While both emission lines were modeled independently, they arise from the same AGN and should agree on some parameters while possibly differing for others.   Comparing and contrasting the results for H$\beta$ and \ion{He}{2} in the context of other studies may thus shed additional light on the Seyfert nucleus of NGC\,3783.

\subsection{Black Hole Mass}

The black hole mass of NGC\,3783 is expected to be the same for both H$\beta$ and \ion{He}{2}.  And indeed, we see that there is significant overlap between the two in the top left panel of Figure~\ref{fig:modelpars}.  We investigated the joint inference on the black hole mass following the method described by \citet{williams20}.  We first approximated the posterior probability distribution functions of each with a Gaussian kernel density estimate and then multiplied them together.  The result is shown in Figure~\ref{fig:mbh} and gives $\log_{10} (M_{\rm BH}/M_{\odot})=7.45^{+0.19}_{-0.11}$, or \mbh$=2.82^{+1.55}_{-0.63}\times10^7$\,\msun.  This is consistent with the simple reverberation constraint on the mass, $M_{\rm BH} = 2.34^{+0.43}_{-0.43} \times 10^7$\,\msun, or $\log_{10} (M_{\rm BH}/M_{\odot})=7.37^{+0.07}_{-0.09}$, which is based on the mean H$\beta$ time delay and line width and an assumed scale factor of $\langle f \rangle=4.82$.  We note that the uncertainties quoted for the simple mass constraint include only the measurement uncertainties on the time delay and line width, and do not include other potential uncertainties such as the object-to-object variation in the scale factor.

With a black hole mass constraint from the BLR models, we can infer a specific value of $f=6.0^{+3.5}_{-1.8}$ for NGC\,3783 using the mean time delay and line width for H$\beta$.  Previous investigations \citep{pancoast14b,grier17,williams18} have found that $f$ scales most strongly with the inclination of the system, as expected because only the line of sight velocity component is measured.  NGC\,3783 seems to follow the same trend that has previously been seen for other Seyferts, as the inclination angle constrained by the models together with the linear regression results of \citet{williams18} predict $f=6^{+16}_{-4}$, or $\log_{10} (f)=0.75^{+0.59}_{-0.61}$.  Thus, the good agreement between our mass constraint and the simple reverberation constraint arises from the inclination of NGC\,3783 being close to the mean inclination value for the sample of local Seyferts, and so having an individual $f$ factor that is similar to the population average.

We can also investigate any changes to the distributions of model parameters that may arise from selecting only those models that agree with the joint H$\beta$ and \ion{He}{2} constraint on \mbh.  Figure~\ref{fig:hb_mbh_weight} shows the constraints on the model parameters for H$\beta$ before and after selecting only those models that agree with the joint \mbh\ constraint.  The results are quite similar, which is unsurprising since the H$\beta$ models provided the strongest initial constraint on \mbh.  Figure~\ref{fig:he2_mbh_weight} shows the same but for \ion{He}{2}.  In this case, we find that models that favor the joint constraint on \mbh\ also favor a slightly larger radius, which makes sense since the joint constraint on \mbh\ was at the upper end of the \mbh\ mass distribution  for \ion{He}{2}, and also favors an even smaller fraction ($\sim15$\%) of bound near-circular orbits with the rest of the orbits outflowing.  No additional changes are seen in the distributions of the model parameters if we similarly constrain the inclination angle in addition to \mbh, in which case we find a joint constraint of $\theta_i=18.2^{+3.6}_{-5.5}$\,degrees.

\subsection{Geometry and Kinematics}

\begin{figure}
    \epsscale{1.17} 
    \plotone{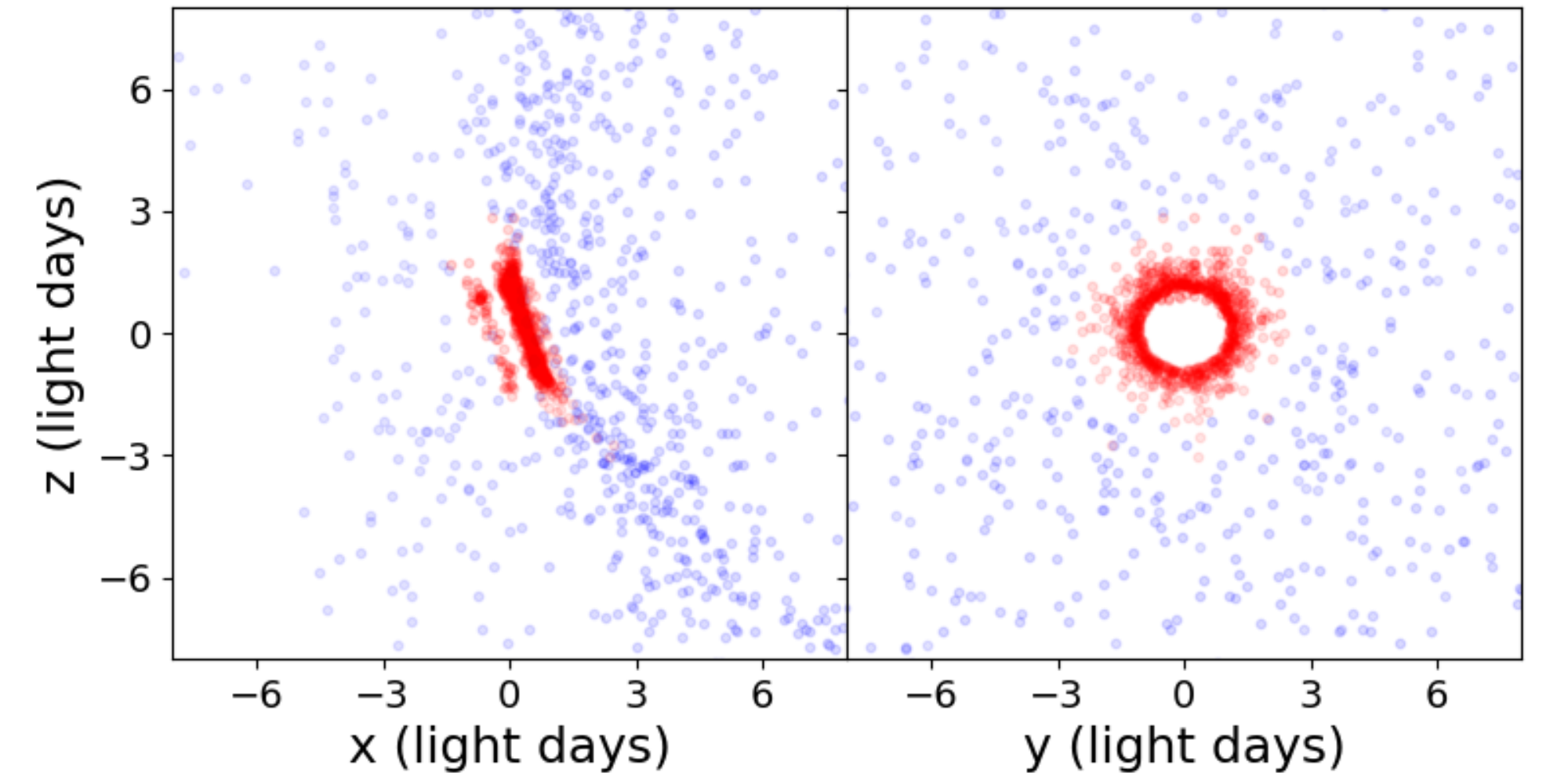}
    \caption{
    Combined representative geometric models for the H$\beta$ response (blue) and \ion{He}{2} response (red) in the broad line region of NGC\,3783.  The left panel is oriented edge on, with an Earth-based observer on the +x axis, while the right panel shows the Earth-based observer's view.  The transparency of each point represents the relative response of the gas to continuum fluctuations at each position, with more opaque points responsible for a stronger response.
    }
    \label{fig:hbhe2clouds}
\end{figure}

The similarities between the inclinations and opening angle constraints for H$\beta$ and \ion{He}{2} support the interpretation that both emission lines are probing different regions of the same thick disk of gas.  While the median values of the opening angles might suggest that the H$\beta$ emitting region is more ``puffed up'' than the \ion{He}{2} emitting region, as might be expected for a bowl-shaped model of the BLR like that proposed by \citet{goad12}, the large uncertainties on the \ion{He}{2} opening angle mean that the two values formally agree.  As expected from the differences in their mean time delays reported by \citet{bentz21a}, the \ion{He}{2}-emitting region is significantly more compact and close to the central ionizing source than the H$\beta$-emitting region, demonstrating clear ionization stratification within the BLR  (e.g., \citealt{peterson93} and references therein).  There is little to no overlap between the two, with $r_{\rm min}=3.25^{+1.13}_{-1.54}$\,light~days for H$\beta$ compared to $r_{\rm mean}=1.40^{+0.31}_{-0.42}$\,light~days and $\sigma_{\rm r}=0.17^{+0.34}_{-0.13}$\,light~days for \ion{He}{2} (see Figure~\ref{fig:hbhe2clouds}).

The two regions of the BLR appear, however, to be dominated by different kinematics.  In the case of the H$\beta$ emitting region, the kinematics are dominated by near-circular orbits with some infall, whereas the \ion{He}{2} emitting region is dominated by outflow.  Repeated studies of the same AGN, such as NGC\,5548 \citep{pancoast14b,williams20}, have found that the best-fit kinematics can change from one reverberation dataset to another, so the different kinematics may not be indicating structural differences between the inner and outer BLR in NGC\,3783, but rather transient effects (``weather'').  On the other hand, \citet{korista04} find that photoionization models predict \ion{He}{2} $\lambda 4686$ is preferentially emitted in the presence of an ionizing photon flux that is $\sim 300$ times stronger than H$\beta$.  While H$\beta$-emitting gas in the BLR has been shown to be fairly stable against radiation pressure \citep{netzer10}, \ion{He}{2} is preferentially emitted from lower density gas \citep{korista04} and may be more susceptible to radiation pressure forces.  

It may be that a combination of weather and photoionization physics  explains the difference in kinematics between H$\beta$ and \ion{He}{2}.  NGC\,3783 has demonstrated possible evidence for changes in the structure of the BLR in the recent past.  \citet{kriss19} obtained UV spectra of NGC\,3783 shortly after the discovery of a strong soft X-ray obscuring event was detected in 2016.  They interpret changes in the UV broad emission lines of NGC\,3783 together with the appearance of new broad absorption lines as evidence that the BLR scale height may have collapsed following a period of low ionizing luminosity that began in 2013 and continued to 2016.  By late 2016, the luminosity had increased significantly and remained high through at least January 2018 \citep{kaastra18}, and could thus begin to drive changes in the structure of the BLR on the dynamical timescale ($\sim0.3$\,years at a BLR radius of 2.0\,light~days, or 3\,years at a radius of 10\,light~days).  The luminosity of NGC\,3783 between early 2018 and early 2020, when our observing campaign began, is unknown, but the BLR may have still been in the process of recovering from the extended low-luminosity period observed in $2013-2016$.   And indeed, a rough comparison of the broad H$\beta$ profile in 2020 with the profiles observed in 2011 and 2016 (Fig.~18; \citealt{kriss19}) suggests that much of the flux deficit observed in the line core in 2016 has filled in, although the line profile has not fully returned to its 2011 state.  Further multiwavelength monitoring coupled with  velocity-resolved reverberation analyses could help to inform our understanding of structural changes in the BLR as a result of large changes in the ionizing luminosity.

Several studies of NGC\,3783 have focused on attempts to model the accretion disk using the Fe K$\alpha$ emission line or the continuum emission \citep{brenneman11,patrick11,capellupo17} and have found similar relatively face-on inclinations for the inner accretion disk, even when they disagree on other components of the models (such as the black hole spin).  A similar inclination angle has also been found by modeling the three-dimensional structure of the spatially-resolved narrow line region on parsec scales \citep{fischer13}. The consistency in inclination angles from the innermost regions of the accretion disk through the broad line region and the outermost narrow line region suggests that the spin axis of the central black hole has been stable for quite some time.  With no evidence for large torques on the spin, and with the black hole spin axis apparently matching the rotation axis of this relatively face-on galaxy, the recent evolution of the supermassive black hole appears to be dominated by secular processes that are aligned with the disk of the galaxy.  

The best-fit models that we find for H$\beta$ also agree well with recent interferometry results for NGC\,3783 from GRAVITY \citep{gravity21}, in which measurements of the broad Br$\gamma$ emission are best described by a rotating thick disk inclined at $\sim 20^{\circ}$ to our line of sight and surrounding a black hole with $\log_{10} (M_{\rm BH}/M_{\odot})=7.68^{+0.45}_{-0.43}$.  Additionally, the radial extent of the Br$\gamma$-emitting region ($r_{\rm mean} = 16^{+12}_{-5}$\,light~days assuming $D=38.5$\,Mpc) is in good agreement with the radial extent of the H$\beta$ emitting region ($r_{\rm mean}=11.4^{+1.1}_{-1.1}$\,light~days;  see Table~\ref{tab:modelpars}).  A joint analysis of the GRAVITY observations with the continuum and integrated H$\beta$ light curves from \citet{bentz21a} confirms and improves upon the results, with $M_{\rm BH}=2.54^{+0.90}_{-0.72}\times10^7$\,\msun, or $\log_{10} (M_{\rm BH}/M_{\odot})=7.40^{+0.13}_{-0.14}$, and $r_{\rm median}=16.2^{+2.8}_{-1.8}$\,light~days \citep{gravity21b}. The black hole mass is still in excellent agreement with our findings, while the stronger constraints on the BLR radius in the joint analysis are somewhat in tension with the size of the BLR reported here ($r_{\rm median}=10.07^{+1.10}_{-1.21}$\,light~days). It is important to recognize that the GRAVITY results depend on the distance to NGC\,3783 which is somewhat uncertain (recent studies suggest values of $35-50$\,Mpc; \citealt{kourkchi20,robinson21}), reverberation mapping measures a responsivity-weighted radius while interferometry measures a flux-weighted radius, and photoionization effects (which are ignored in both our models and those employed in the analysis of the GRAVITY data) are known to cause different reverberation time delays for different Hydrogen recombination lines (e.g., \citealt{bentz10a}).  Despite these complicating factors, the good agreement between the results lends additional confidence to both.  
Future work will investigate the joint constraints that may be derived from an analysis of the velocity-resolved reverberation data that we have presented here in tandem with the GRAVITY observations.

\section{Summary}

We have modeled the full velocity-resolved response of the broad H$\beta$ and \ion{He}{2} emission lines in NGC\,3783.  The results give a black hole mass constraint that is independent of any scaling factor, and a joint analysis of the results for the two emission lines prefers \mbh$=2.82^{+1.55}_{-0.63}\times10^7$\,\msun.  The geometry of the BLR is found to be a thick disk that is close to face on ($\theta_i\approx 18^{\circ}$) and exhibiting clear ionization stratification, with H$\beta$ arising from an extended region of $~\sim 3-20$\,light-days, while \ion{He}{2} arises from a significantly more compact and centralized region of $1-2$\,light~days.  The kinematics of the outer BLR probed by H$\beta$ are dominated by near-circular orbits with a contribution from infall, whereas the kinematics of the inner BLR probed by \ion{He}{2} are dominated by an unbound outflow.  Given the recent history of a deficit of ionizing radiation in NGC\,3783 that was observed from $2013-2016$, and the hypothesis that the BLR height collapsed as a result, it is possible that we may be seeing the BLR undergoing structural changes as it recovers.

\begin{acknowledgements}
We thank the anonymous referee for suggestions that improved the presentation of this work.
We also thank Kate Grier for helpful conversations about CARAMEL.
MCB gratefully acknowledges support from the NSF through grant AST-2009230.
TT and PRW acknowledge support by the Packard Foundation through a Packard Research Fellowship to TT and from NSF through grant NSF-AST-1907208. PRW acknowledges support from the UCLA graduate division through a Dissertation Year Fellowship.
\end{acknowledgements}


\facility{LCOGT}

\software{ULySS \citep{koleva11}, CARAMEL \citep{pancoast14a} }


\begin{thebibliography}{}
\expandafter\ifx\csname natexlab\endcsname\relax\def\natexlab#1{#1}\fi
\providecommand{\url}[1]{\href{#1}{#1}}
\providecommand{\dodoi}[1]{doi:~\href{http://doi.org/#1}{\nolinkurl{#1}}}
\providecommand{\doeprint}[1]{\href{http://ascl.net/#1}{\nolinkurl{http://ascl.net/#1}}}
\providecommand{\doarXiv}[1]{\href{https://arxiv.org/abs/#1}{\nolinkurl{https://arxiv.org/abs/#1}}}

\bibitem[{{Alard}(2000)}]{alard00}
{Alard}, C. 2000, \aaps, 144, 363, \dodoi{10.1051/aas:2000214}

\bibitem[{{Alard} \& {Lupton}(1998)}]{alard98}
{Alard}, C., \& {Lupton}, R.~H. 1998, \apj, 503, 325, \dodoi{10.1086/305984}

\bibitem[{{Anderson} {et~al.}(2021){Anderson}, {Baron}, \&
  {Bentz}}]{anderson21}
{Anderson}, M.~D., {Baron}, F., \& {Bentz}, M.~C. 2021, \mnras,
  \dodoi{10.1093/mnras/stab1394}

\bibitem[{{Bahcall} {et~al.}(1972){Bahcall}, {Kozlovsky}, \&
  {Salpeter}}]{bahcall72}
{Bahcall}, J.~N., {Kozlovsky}, B.-Z., \& {Salpeter}, E.~E. 1972, \apj, 171,
  467, \dodoi{10.1086/151300}

\bibitem[{{Batiste} {et~al.}(2017){Batiste}, {Bentz}, {Raimundo},
  {Vestergaard}, \& {Onken}}]{batiste17b}
{Batiste}, M., {Bentz}, M.~C., {Raimundo}, S.~I., {Vestergaard}, M., \&
  {Onken}, C.~A. 2017, \apjl, 838, L10, \dodoi{10.3847/2041-8213/aa6571}

\bibitem[{{Bentz} {et~al.}(2010{\natexlab{a}}){Bentz}, {Horne}, {Barth},
  {et~al.}}]{bentz10b}
{Bentz}, M.~C., {Horne}, K., {Barth}, A.~J., {et~al.} 2010{\natexlab{a}},
  \apjl, 720, L46, \dodoi{10.1088/2041-8205/720/1/L46}

\bibitem[{{Bentz} {et~al.}(2021){Bentz}, {Street}, {Onken}, \&
  {Valluri}}]{bentz21a}
{Bentz}, M.~C., {Street}, R., {Onken}, C.~A., \& {Valluri}, M. 2021, \apj, 906,
  50, \dodoi{10.3847/1538-4357/abccd4}

\bibitem[{{Bentz} {et~al.}(2009){Bentz}, {Walsh}, {Barth}, {et~al.}}]{bentz09c}
{Bentz}, M.~C., {Walsh}, J.~L., {Barth}, A.~J., {et~al.} 2009, \apj, 705, 199,
  \dodoi{10.1088/0004-637X/705/1/199}

\bibitem[{{Bentz} {et~al.}(2010{\natexlab{b}}){Bentz}, {Walsh}, {Barth},
  {et~al.}}]{bentz10a}
---. 2010{\natexlab{b}}, \apj, 716, 993, \dodoi{10.1088/0004-637X/716/2/993}

\bibitem[{{Bentz} {et~al.}(2008)}]{bentz08}
{Bentz}, M.~C., {et~al.} 2008, \apjl, 689, L21, \dodoi{10.1086/595719}

\bibitem[{{Blandford} \& {McKee}(1982)}]{blandford82}
{Blandford}, R.~D., \& {McKee}, C.~F. 1982, \apj, 255, 419

\bibitem[{{Brenneman} {et~al.}(2011){Brenneman}, {Reynolds}, {Nowak}, {Reis},
  {Trippe}, {Fabian}, {Iwasawa}, {Lee}, {Miller}, {Mushotzky}, {Nandra}, \&
  {Volonteri}}]{brenneman11}
{Brenneman}, L.~W., {Reynolds}, C.~S., {Nowak}, M.~A., {et~al.} 2011, \apj,
  736, 103, \dodoi{10.1088/0004-637X/736/2/103}

\bibitem[{Brewer \& Foreman-Mackey(2018)}]{brewer18}
Brewer, B.~J., \& Foreman-Mackey, D. 2018, Journal of Statistical Software,
  Articles, 86, 1, \dodoi{10.18637/jss.v086.i07}

\bibitem[{Cackett {et~al.}(2021)Cackett, Bentz, \& Kara}]{cackett21}
Cackett, E.~M., Bentz, M.~C., \& Kara, E. 2021, iScience, 24, 102557,
  \dodoi{https://doi.org/10.1016/j.isci.2021.102557}

\bibitem[{{Capellupo} {et~al.}(2017){Capellupo}, {Wafflard-Fernandez}, \&
  {Haggard}}]{capellupo17}
{Capellupo}, D.~M., {Wafflard-Fernandez}, G., \& {Haggard}, D. 2017, \apjl,
  836, L8, \dodoi{10.3847/2041-8213/aa5cac}

\bibitem[{{Denney} {et~al.}(2009){Denney}, {Peterson}, {Pogge}, {Adair},
  {Atlee}, {Au-Yong}, {Bentz}, {Bird}, {Brokofsky}, {Chisholm}, {Comins},
  {Dietrich}, {Doroshenko}, {Eastman}, {Efimov}, {Ewald}, {Ferbey}, {Gaskell},
  {Hedrick}, {Jackson}, {Klimanov}, {Klimek}, {Kruse}, {Lad{\'e}route}, {Lamb},
  {Leighly}, {Minezaki}, {Nazarov}, {Onken}, {Petersen}, {Peterson},
  {Poindexter}, {Sakata}, {Schlesinger}, {Sergeev}, {Skolski}, {Stieglitz},
  {Tobin}, {Unterborn}, {Vestergaard}, {Watkins}, {Watson}, \&
  {Yoshii}}]{denney09c}
{Denney}, K.~D., {Peterson}, B.~M., {Pogge}, R.~W., {et~al.} 2009, \apjl, 704,
  L80, \dodoi{10.1088/0004-637X/704/2/L80}

\bibitem[{{Event Horizon Telescope Collaboration} {et~al.}(2019){Event Horizon
  Telescope Collaboration}, {Akiyama}, {Alberdi}, {et~al.}}]{eht19}
{Event Horizon Telescope Collaboration}, {Akiyama}, K., {Alberdi}, A., {et~al.}
  2019, \apjl, 875, L6, \dodoi{10.3847/2041-8213/ab1141}

\bibitem[{{Ferrarese} \& {Merritt}(2000)}]{ferrarese00}
{Ferrarese}, L., \& {Merritt}, D. 2000, \apjl, 539, L9

\bibitem[{{Fischer} {et~al.}(2013){Fischer}, {Crenshaw}, {Kraemer}, \&
  {Schmitt}}]{fischer13}
{Fischer}, T.~C., {Crenshaw}, D.~M., {Kraemer}, S.~B., \& {Schmitt}, H.~R.
  2013, \apjs, 209, 1, \dodoi{10.1088/0067-0049/209/1/1}

\bibitem[{{Gebhardt} {et~al.}(2000){Gebhardt}, {Bender}, {Bower},
  {et~al.}}]{gebhardt00}
{Gebhardt}, K., {Bender}, R., {Bower}, G., {et~al.} 2000, \apjl, 539, L13

\bibitem[{{Genzel} {et~al.}(2000){Genzel}, {Pichon}, {Eckart}, {Gerhard}, \&
  {Ott}}]{genzel00}
{Genzel}, R., {Pichon}, C., {Eckart}, A., {Gerhard}, O.~E., \& {Ott}, T. 2000,
  \mnras, 317, 348, \dodoi{10.1046/j.1365-8711.2000.03582.x}

\bibitem[{{Ghez} {et~al.}(2000){Ghez}, {Morris}, {Becklin}, {Tanner}, \&
  {Kremenek}}]{ghez00}
{Ghez}, A.~M., {Morris}, M., {Becklin}, E.~E., {Tanner}, A., \& {Kremenek}, T.
  2000, \nat, 407, 349, \dodoi{10.1038/407349a}

\bibitem[{{Ghez} {et~al.}(2008){Ghez}, {Salim}, {Weinberg}, {Lu}, {Do}, {Dunn},
  {Matthews}, {Morris}, {Yelda}, {Becklin}, {Kremenek}, {Milosavljevic}, \&
  {Naiman}}]{ghez08}
{Ghez}, A.~M., {Salim}, S., {Weinberg}, N.~N., {et~al.} 2008, \apj, 689, 1044,
  \dodoi{10.1086/592738}

\bibitem[{{Goad} {et~al.}(2012){Goad}, {Korista}, \& {Ruff}}]{goad12}
{Goad}, M.~R., {Korista}, K.~T., \& {Ruff}, A.~J. 2012, \mnras, 426, 3086,
  \dodoi{10.1111/j.1365-2966.2012.21808.x}

\bibitem[{{Gravity Collaboration} {et~al.}(2021{\natexlab{a}}){Gravity
  Collaboration}, {Amorim}, {Baub{\"o}ck}, {Brandner}, {Bolzer}, {Cl{\'e}net},
  {Davies}, {de Zeeuw}, {Dexter}, {Drescher}, {Eckart}, {Eisenhauer},
  {F{\"o}rster Schreiber}, {Gao}, {Garcia}, {Genzel}, {Gillessen}, {Gratadour},
  {H{\"o}nig}, {Kaltenbrunner}, {Kishimoto}, {Lacour}, {Lutz}, {Millour},
  {Netzer}, {Ott}, {Paumard}, {Perraut}, {Perrin}, {Peterson}, {Petrucci},
  {Pfuhl}, {Prieto}, {Rouan}, {Sanchez-Bermudez}, {Shangguan}, {Shimizu},
  {Schartmann}, {Stadler}, {Sternberg}, {Straub}, {Straubmeier}, {Sturm},
  {Tacconi}, {Tristram}, {Vermot}, {von Fellenberg}, {Waisberg}, {Widmann}, \&
  {Woillez}}]{gravity21}
{Gravity Collaboration}, {Amorim}, A., {Baub{\"o}ck}, M., {et~al.}
  2021{\natexlab{a}}, \aap, 648, A117, \dodoi{10.1051/0004-6361/202040061}

\bibitem[{{Gravity Collaboration} {et~al.}(2021{\natexlab{b}}){Gravity
  Collaboration}, {Amorim}, {Baub{\"o}ck}, {Bentz}, {Brandner}, {Bolzer},
  {Cl{\'e}net}, {Davies}, {de Zeeuw}, {Dexter}, {Drescher}, {Eckart},
  {Eisenhauer}, {F{\"o}rster Schreiber}, {Garcia}, {Genzel}, {Gillessen},
  {Gratadour}, {H{\"o}nig}, {Kaltenbrunner}, {Kishimoto}, {Lacour}, {Lutz},
  {Millour}, {Netzer}, {Onken}, {Ott}, {Paumard}, {Perraut}, {Perrin},
  {Peterson}, {Petrucci}, {Pfuhl}, {Prieto}, {Rouan}, {Shangguan}, {Shimizu},
  {Stadler}, {Sternberg}, {Straub}, {Straubmeier}, {Street}, {Sturm},
  {Tacconi}, {Tristram}, {Valluri}, {Vermot}, {von Fellenberg}, {Widmann}, \&
  {Woillez}}]{gravity21b}
---. 2021{\natexlab{b}}, \aap, submitted

\bibitem[{{Grier} {et~al.}(2013){Grier}, {Martini}, {Watson},
  {et~al.}}]{grier13}
{Grier}, C.~J., {Martini}, P., {Watson}, L.~C., {et~al.} 2013, \apj, 773, 90,
  \dodoi{10.1088/0004-637X/773/2/90}

\bibitem[{{Grier} {et~al.}(2017){Grier}, {Pancoast}, {Barth}, {Fausnaugh},
  {Brewer}, {Treu}, \& {Peterson}}]{grier17}
{Grier}, C.~J., {Pancoast}, A., {Barth}, A.~J., {et~al.} 2017, \apj, 849, 146,
  \dodoi{10.3847/1538-4357/aa901b}

\bibitem[{{Grier} {et~al.}(2012){Grier}, {Peterson}, {Pogge},
  {et~al.}}]{grier12b}
{Grier}, C.~J., {Peterson}, B.~M., {Pogge}, R.~W., {et~al.} 2012, \apj, 755,
  60, \dodoi{10.1088/0004-637X/755/1/60}

\bibitem[{{G{\"u}ltekin} {et~al.}(2009){G{\"u}ltekin}, {Richstone}, {Gebhardt},
  {et~al.}}]{gultekin09}
{G{\"u}ltekin}, K., {Richstone}, D.~O., {Gebhardt}, K., {et~al.} 2009, \apj,
  698, 198, \dodoi{10.1088/0004-637X/698/1/198}

\bibitem[{{Horne}(1994)}]{horne94}
{Horne}, K. 1994, in Astronomical Society of the Pacific Conference Series,
  Vol.~69, Reverberation Mapping of the Broad-Line Region in Active Galactic
  Nuclei, ed. P.~M. {Gondhalekar}, K.~{Horne}, \& B.~M. {Peterson}, 23

\bibitem[{{Horne} {et~al.}(2021){Horne}, {De Rosa}, {Peterson}, {Barth}, {Ely},
  {Fausnaugh}, {Kriss}, {Pei}, {Bentz}, {Cackett}, {Edelson}, {Eracleous},
  {Goad}, {Grier}, {Kaastra}, {Kochanek}, {Krongold}, {Mathur}, {Netzer},
  {Proga}, {Tejos}, {Vestergaard}, {Villforth}, {Adams}, {Anderson},
  {Ar{\'e}valo}, {Beatty}, {Bennert}, {Bigley}, {Bisogni}, {Borman}, {Boroson},
  {Bottorff}, {Brandt}, {Breeveld}, {Brotherton}, {Brown}, {Brown}, {Canalizo},
  {Carini}, {Clubb}, {Comerford}, {Corsini}, {Crenshaw}, {Croft}, {Croxall},
  {Dalla Bont{\`a}}, {Deason}, {Dehghanian}, {De Lorenzo-C{\'a}ceres},
  {Denney}, {Dietrich}, {Done}, {Efimova}, {Evans}, {Ferland}, {Filippenko},
  {Flatland}, {Fox}, {Gardner}, {Gates}, {Gehrels}, {Geier}, {Gelbord},
  {Gonzalez}, {Gorjian}, {Greene}, {Grupe}, {Gupta}, {Hall}, {Henderson},
  {Hicks}, {Holmbeck}, {Holoien}, {Hutchison}, {Im}, {Jensen}, {Johnson},
  {Joner}, {Jones}, {Kaspi}, {Kelly}, {Kennea}, {Kim}, {Kim}, {Kim}, {King},
  {Klimanov}, {Korista}, {Lau}, {Lee}, {Leonard}, {Li}, {Lira}, {Lochhaas},
  {Ma}, {MacInnis}, {Malkan}, {Manne-Nicholas}, {Mauerhan}, {McGurk},
  {McHardy}, {Montuori}, {Morelli}, {Mosquera}, {Mudd},
  {M{\"u}ller-S{\'a}nchez}, {Nazarov}, {Norris}, {Nousek}, {Nguyen}, {Ochner},
  {Okhmat}, {Pancoast}, {Papadakis}, {Parks}, {Penny}, {Pizzella}, {Pogge},
  {Poleski}, {Pott}, {Rafter}, {Rix}, {Runnoe}, {Saylor}, {Schimoia},
  {Schn{\"u}lle}, {Scott}, {Sergeev}, {Shappee}, {Shivvers}, {Siegel},
  {Simonian}, {Siviero}, {Skielboe}, {Somers}, {Spencer}, {Starkey}, {Stevens},
  {Sung}, {Tayar}, {Treu}, {Turner}, {Uttley}, {Van Saders}, {Vican},
  {Villanueva}, {Weiss}, {Woo}, {Yan}, {Young}, {Yuk}, {Zheng}, {Zhu}, \&
  {Zu}}]{horne21}
{Horne}, K., {De Rosa}, G., {Peterson}, B.~M., {et~al.} 2021, \apj, 907, 76,
  \dodoi{10.3847/1538-4357/abce60}

\bibitem[{{Kaastra} {et~al.}(2018){Kaastra}, {Mehdipour}, {Behar}, {Bianchi},
  {Branduardi-Raymont}, {Brenneman}, {Cappi}, {Costantini}, {De Marco}, {di
  Gesu}, {Ebrero}, {Kriss}, {Mao}, {Peretz}, {Petrucci}, {Ponti}, \&
  {Walton}}]{kaastra18}
{Kaastra}, J.~S., {Mehdipour}, M., {Behar}, E., {et~al.} 2018, \aap, 619, A112,
  \dodoi{10.1051/0004-6361/201832629}

\bibitem[{{Koleva} {et~al.}(2009){Koleva}, {Prugniel}, {Bouchard}, \&
  {Wu}}]{koleva09}
{Koleva}, M., {Prugniel}, P., {Bouchard}, A., \& {Wu}, Y. 2009, \aap, 501,
  1269, \dodoi{10.1051/0004-6361/200811467}

\bibitem[{{Koleva} {et~al.}(2011){Koleva}, {Prugniel}, {Bouchard}, \&
  {Wu}}]{koleva11}
---. 2011, {ULySS: A Full Spectrum Fitting Package}.
\newblock \doeprint{1104.007}

\bibitem[{{Korista} \& {Goad}(2004)}]{korista04}
{Korista}, K.~T., \& {Goad}, M.~R. 2004, \apj, 606, 749

\bibitem[{{Kormendy} \& {Ho}(2013)}]{kormendy13}
{Kormendy}, J., \& {Ho}, L.~C. 2013, \araa, 51, 511,
  \dodoi{10.1146/annurev-astro-082708-101811}

\bibitem[{{Kourkchi} {et~al.}(2020){Kourkchi}, {Courtois}, {Graziani},
  {Hoffman}, {Pomar{\`e}de}, {Shaya}, \& {Tully}}]{kourkchi20}
{Kourkchi}, E., {Courtois}, H.~M., {Graziani}, R., {et~al.} 2020, \aj, 159, 67,
  \dodoi{10.3847/1538-3881/ab620e}

\bibitem[{{Kriss} {et~al.}(2019){Kriss}, {Mehdipour}, {Kaastra}, {Rau},
  {Bodensteiner}, {Plesha}, {Arav}, {Behar}, {Bianchi}, {Branduardi-Raymont},
  {Cappi}, {Costantini}, {De Marco}, {Di Gesu}, {Ebrero}, {Kaspi}, {Mao},
  {Middei}, {Miller}, {Paltani}, {Peretz}, {Peterson}, {Petrucci}, {Ponti},
  {Ursini}, {Walton}, \& {Xu}}]{kriss19}
{Kriss}, G.~A., {Mehdipour}, M., {Kaastra}, J.~S., {et~al.} 2019, \aap, 621,
  A12, \dodoi{10.1051/0004-6361/201834326}

\bibitem[{{Magorrian} {et~al.}(1998){Magorrian}, {Tremaine}, {Richstone},
  {Bender}, {Bower}, {Dressler}, {Faber}, {Gebhardt}, {Green}, {Grillmair},
  {Kormendy}, \& {Lauer}}]{magorrian98}
{Magorrian}, J., {Tremaine}, S., {Richstone}, D., {et~al.} 1998, \aj, 115, 2285

\bibitem[{{Michell}(1784)}]{michell1784}
{Michell}, J. 1784, Philosophical Transactions of the Royal Society of London
  Series I, 74, 35

\bibitem[{{Netzer} \& {Marziani}(2010)}]{netzer10}
{Netzer}, H., \& {Marziani}, P. 2010, \apj, 724, 318,
  \dodoi{10.1088/0004-637X/724/1/318}

\bibitem[{{Onken} {et~al.}(2004){Onken}, {Ferrarese}, {Merritt}, {Peterson},
  {Pogge}, {Vestergaard}, \& {Wandel}}]{onken04}
{Onken}, C.~A., {Ferrarese}, L., {Merritt}, D., {et~al.} 2004, \apj, 615, 645,
  \dodoi{10.1086/424655}

\bibitem[{{Pancoast} {et~al.}(2011){Pancoast}, {Brewer}, \&
  {Treu}}]{pancoast11}
{Pancoast}, A., {Brewer}, B.~J., \& {Treu}, T. 2011, \apj, 730, 139,
  \dodoi{10.1088/0004-637X/730/2/139}

\bibitem[{{Pancoast} {et~al.}(2014{\natexlab{a}}){Pancoast}, {Brewer}, \&
  {Treu}}]{pancoast14a}
---. 2014{\natexlab{a}}, \mnras, 445, 3055, \dodoi{10.1093/mnras/stu1809}

\bibitem[{{Pancoast} {et~al.}(2014{\natexlab{b}}){Pancoast}, {Brewer}, {Treu},
  {Park}, {Barth}, {Bentz}, \& {Woo}}]{pancoast14b}
{Pancoast}, A., {Brewer}, B.~J., {Treu}, T., {et~al.} 2014{\natexlab{b}},
  \mnras, 445, 3073, \dodoi{10.1093/mnras/stu1419}

\bibitem[{{Park} {et~al.}(2012){Park}, {Kelly}, {Woo}, \& {Treu}}]{park12}
{Park}, D., {Kelly}, B.~C., {Woo}, J.-H., \& {Treu}, T. 2012, \apjs, 203, 6,
  \dodoi{10.1088/0067-0049/203/1/6}

\bibitem[{{Patrick} {et~al.}(2011){Patrick}, {Reeves}, {Lobban}, {Porquet}, \&
  {Markowitz}}]{patrick11}
{Patrick}, A.~R., {Reeves}, J.~N., {Lobban}, A.~P., {Porquet}, D., \&
  {Markowitz}, A.~G. 2011, \mnras, 416, 2725,
  \dodoi{10.1111/j.1365-2966.2011.19224.x}

\bibitem[{{Peterson}(1993)}]{peterson93}
{Peterson}, B.~M. 1993, \pasp, 105, 247

\bibitem[{{Peterson} {et~al.}(2013){Peterson}, {Denney}, {De Rosa},
  {et~al.}}]{peterson13}
{Peterson}, B.~M., {Denney}, K.~D., {De Rosa}, G., {et~al.} 2013, \apj, 779,
  109, \dodoi{10.1088/0004-637X/779/2/109}

\bibitem[{{Peterson} \& {Wandel}(1999)}]{peterson99}
{Peterson}, B.~M., \& {Wandel}, A. 1999, \apjl, 521, L95

\bibitem[{{Peterson} \& {Wandel}(2000)}]{peterson00a}
---. 2000, \apjl, 540, L13

\bibitem[{{Robinson} {et~al.}(2021){Robinson}, {Bentz}, {Courtois}, {Johnson},
  {Crenshaw}, {Meena}, {Polack}, {Silverstein}, \& {Chen}}]{robinson21}
{Robinson}, J.~H., {Bentz}, M.~C., {Courtois}, H.~M., {et~al.} 2021, \apj, 912,
  160, \dodoi{10.3847/1538-4357/abedaa}

\bibitem[{{Skielboe} {et~al.}(2015){Skielboe}, {Pancoast}, {Treu}, {Park},
  {Barth}, \& {Bentz}}]{skielboe15}
{Skielboe}, A., {Pancoast}, A., {Treu}, T., {et~al.} 2015, \mnras, 454, 144,
  \dodoi{10.1093/mnras/stv1917}

\bibitem[{{Vazdekis} {et~al.}(2010){Vazdekis}, {S{\'a}nchez-Bl{\'a}zquez},
  {Falc{\'o}n-Barroso}, {Cenarro}, {Beasley}, {Cardiel}, {Gorgas}, \&
  {Peletier}}]{vazdekis10}
{Vazdekis}, A., {S{\'a}nchez-Bl{\'a}zquez}, P., {Falc{\'o}n-Barroso}, J.,
  {et~al.} 2010, \mnras, 404, 1639, \dodoi{10.1111/j.1365-2966.2010.16407.x}

\bibitem[{{Williams} {et~al.}(2018){Williams}, {Pancoast}, {Treu},
  {et~al.}}]{williams18}
{Williams}, P.~R., {Pancoast}, A., {Treu}, T., {et~al.} 2018, \apj, 866, 75,
  \dodoi{10.3847/1538-4357/aae086}

\bibitem[{{Williams} {et~al.}(2020){Williams}, {Pancoast}, {Treu},
  {et~al.}}]{williams20}
---. 2020, \apj, 902, 74, \dodoi{10.3847/1538-4357/abbad7}

\end{thebibliography}
\bibliographystyle{aasjournal}

\end{document}